\documentclass[aps,pre,twocolumn,superscriptaddress,amsmath,amsfonts,showpacs]{revtex4}
\usepackage{graphicx}
\usepackage{subfigure}
\usepackage{color}
\newcommand{\pap}{\psi_{\rm app}}

\begin{document}

\title{Intrinsic localized modes in parametrically-driven arrays
  of nonlinear resonators}

\author{Eyal Kenig}
\affiliation{Raymond and Beverly Sackler School of Physics and
  Astronomy, Tel Aviv University, Tel Aviv 69978, Israel}
\author{Boris A.~Malomed}
\affiliation{Department of Physical Electronics, School of Electrical
  Engineering, Tel Aviv University, Tel Aviv 69978, Israel}
\author{M.~C.~Cross}
\affiliation{Condensed Matter Physics 114-36, California Institute of
  Technology, Pasadena, California 91125, USA}
\author{Ron Lifshitz}
\email[Corresponding author:\ ]{ronlif@tau.ac.il}
\affiliation{Raymond and Beverly Sackler School of Physics and
  Astronomy, Tel Aviv University, Tel Aviv 69978, Israel}

\date{\today}
\begin{abstract}
  We study intrinsic localized modes (ILMs), or solitons, in arrays of
  parametrically-driven nonlinear resonators with application to
  microelectromechanical and nanoelectromechanical systems (MEMS and
  NEMS). The analysis is performed using an amplitude equation in the
  form of a nonlinear Schr\"odinger equation with a term corresponding
  to nonlinear damping (also known as a forced complex Ginzburg-Landau
  equation), which is derived directly from the underlying equations
  of motion of the coupled resonators, using the method of multiple
  scales. We investigate the creation, stability, and interaction of
  ILMs, show that they can form bound states, and that under certain
  conditions one ILM can split into two. Our findings are confirmed by
  simulations of the underlying equations of motion of the resonators,
  suggesting possible experimental tests of the theory.
\end{abstract}

\maketitle

\section{INTRODUCTION}

The study of collective nonlinear dynamics of coupled
\emph{mechanical} resonators has been regaining attention in recent
years~\cite{LCreview} thanks to advances in fabrication,
transduction, and detection of microelectromechanical and
nanoelectromechanical systems (MEMS and NEMS). Nonlinearity is readily
observed in these systems~\cite{turner98,C00,BR01,%
  blick02,turner02,turner03,yu02,cleland04,erbe00,kozinsky07,%
  demartini07,masmanidis07}, and is even proposed as a way to detect
quantum behavior~\cite{katz07,katz08}. Typical MEMS and NEMS
resonators are characterized by extremely high frequencies---from
hundreds of kHz to a few GHz~\cite{HZMR03,cleland:070501}---and
relatively weak dissipation, with quality factors $Q$ in the range of
$10^{2}-10^{5}$.  For such devices, under external driving conditions,
transients die out rapidly, making it is easy to acquire sufficient
data to characterize the steady-state well. Because the basic physics
of the individual elements is simple, and relevant parameters can
readily be measured or calculated, the equations of motion describing
the system can be established with confidence. This, and the fact that
weak dissipation can be treated perturbatively, are a great
advantage for comparison between theory and experiment.

Current technology enables the fabrication of large arrays, composed
of hundreds or thousands of MEMS and NEMS devices, coupled by
electric, magnetic, or elastic forces. These arrays offer new
possibilities for quantitative studies of nonlinear dynamics in
systems with an intermediate number of degrees of freedom---much
larger than one can deal with in macroscopic experiments, yet much
smaller than one confronts when considering nonlinear aspects of
phonon dynamics in a crystal. Our studies of collective nonlinear
dynamics of MEMS and NEMS were originally motivated by the experiment
of Buks and Roukes~\cite{BR}, in which an array of 67 doubly-clamped
micromechanical gold beams was parametrically excited by modulating
the strength of an externally-controlled electrostatic coupling
between neighboring beams. These studies have led to a quantitative
understanding of the collective response of such an array, providing
explicit bifurcation diagrams that explain the transitions between
different extended modes of the array as the strength and frequency of
the external drive are varied quasistatically~\cite{LC,BCL}. We have
also considered more general issues such as the nonlinear competition
between extended modes, or patterns, of the system---when many such
patterns are simultaneously stable---as the external driving
parameters are changed abruptly or ramped as a function of
time~\cite{kenig}. Furthermore, we have investigated the
synchronization that may occur in coupled arrays of
\emph{non-identical} nonlinear resonators, based on the ability of
nonlinear resonators to tune their frequency by changing their
oscillation amplitude~\cite{sync1,sync2}.

Here we focus on a different type of nonlinear states, namely,
intrinsic localized modes (ILMs), also known as discrete breathers or
lattice solitons~\cite{sievers88,campbell04,Maniadis06}. These
localized states are intrinsic in the sense that they arise from the
inherent nonlinearity of the resonators, rather than from
extrinsically-imposed disorder as in the case of Anderson
localization. ILMs have been observed by Sato \emph{et al.}~\cite{%
  sato03_1,sato03_2,sato04,sato06,sato07,sato08} in driven arrays of
micromechanical resonators. They have also been observed in a wide
range of other physical systems including coupled arrays of Josephson
junctions~\cite{trias,binder00}, coupled optical
waveguides~\cite{eisenberg,eisenberg01,shimshon1}, two-dimensional
nonlinear photonic crystals~\cite{fleischer}, highly-nonlinear atomic
lattices~\cite{swanson}, and
antiferromagnets~\cite{schwarz,satoAFM04}. Thus, the ability to
perform a quantitative comparison between our theory and future
experiments with large arrays of MEMS and NEMS resonators, may have
consequences far beyond the framework of mechanical systems considered
here.

We aim to predict the actual physical parameters, in realistic arrays
of MEMS and NEMS resonators, for which ILMs can form and sustain
themselves. Such predictions may have practical consequences for
actual applications exploiting self-localization to focus energy, and
others that may want to avoid energy focusing, for example in cases
where very large oscillation amplitudes may lead to mechanical
failure.  Although quantitative analysis can be carried out directly
in the framework of the underlying oscillator equations of motion, it
is instructive to formulate the analysis in terms of an amplitude
equation, as done previously for extended modes~\cite{LCreview,BCL,%
  kenig}. This allows one to display the range of stable ILMs on a
reduced diagram, helping to describe the general qualitative behavior
as physical parameters are varied.  In Sec.~\ref{derivation} we
describe the derivation of such a single amplitude equation from the
coupled equations of motion that model an array of nonlinear
resonators. This amplitude equation is obtained in the form of a
parametrically-driven damped nonlinear Schr\"odinger equation with an
additional nonlinear damping term, also known as the forced complex
Ginzburg-Landau equation. In most physical systems, the dissipation of
energy is modeled by a linear damping term. However, it has been
established, both theoretically~\cite{LCreview,LC} and
experimentally~\cite{zaitsev}, that nonlinear damping is important for
correct modeling of certain high-$Q$ nonlinear MEMS and NEMS
resonators.

In Sec.~\ref{witheta} we argue that exact soliton solutions that exist
in the absence of nonlinear damping can be continued to solve the full
amplitude equation, with nonlinear damping. We derive an approximate
analytical expression for these solitons, and then use it to find the
exact soliton solutions numerically. A linear stability analysis of
these soliton solutions follows in Sec.~\ref{stability}. In
Sec.~\ref{flat} we consider the dynamical formation of solitons and
study the effects of nonlinear damping on the modulational instability
of non-zero uniform solutions of the amplitude equation.  In
Sec.~\ref{interactions} we study the interactions between pairs of
solitons; and in Sec.~\ref{splitting} demonstrate the splitting of a
single soliton into two separate ones.  Finally, we show in
Sec.~\ref{complexes} that solitons of the full amplitude equation can
form stable bound states. As emphasized in the concluding remarks in
Sec.~\ref{sec:conclusions}, all the phenomena demonstrated through the
analysis of the amplitude equation are accurately reproduced in
simulations of the underlying equations of motion of the coupled
resonators, and therefore should also be reproducible in actual
experiments with MEMS and NEMS arrays.

\section{DERIVATION OF THE AMPLITUDE EQUATION}
\label{derivation}

Lifshitz and Cross~\cite{LC} modeled the array of coupled
nonlinear resonators that was studied by Buks and
Roukes~\cite{BR} using the equations of motion
\begin{align}
  \label{eom}
  \ddot{u}_{n} & + u_{n} + u_{n}^{3} -
  \frac{1}{2}Q^{-1}(\dot{u}_{n+1} - 2\dot{u}_{n} +
  \dot{u}_{n-1})\nonumber\\
  & + \frac{1}{2}\bigl[D + H\cos(2\omega_{p}t)\bigr](u_{n+1}%
  - 2u_{n} + u_{n-1})\nonumber\\
  & - \frac{1}{2}\hat{\eta}\bigl[(u_{n+1} - u_{n})^{2}(\dot{u}_{n+1} -
  \dot{u}_{n})\nonumber\\
  &  - (u_{n}-u_{n-1})^{2}(\dot{u}_{n} - \dot{u}_{n-1})\bigr] = 0,
\end{align}
where $u_{n}$ describes the deviation of the $n^{th}$ resonator from
its equilibrium position, with $n=1\ldots N$, and fixed boundary
conditions $u_{0}=u_{N+1}=0$. Detailed arguments for the particular
choice of terms introduced into the equations of motion are discussed
in Ref.~\cite{LC}.  The terms include an elastic restoring force with
both linear and cubic contributions (whose coefficients are both
scaled to 1), a dc electrostatic nearest-neighbor coupling term with a
small ac component responsible for the parametric excitation (with
coefficients $D$ and $H$ respectively), and linear as well as cubic
nonlinear dissipation terms. The nonlinear elastic term is positive,
indicating a stiffening of the resonators with increasing
displacement, which is the common situation when using doubly-clamped
beams. Both dissipation terms are taken in the nearest-neighbor form,
which is motivated by the experimental indication that most of the
dissipation originates from the electrostatic interaction between
adjacent beams. Note that the electrostatic attractive force, acting
between neighboring beams, decays with distance, and thus acts to
soften the elastic restoring force. For this reason the sign in front
of the coupling coefficient $D$ is positive, and accordingly the
dispersion curve in the linear regime features a negative slope, or a
negative group velocity.

In more recent implementations~\cite{masmanidis07}, the electric
current damping has been reduced, and the parametric drive is applied
piezoelectrically directly to each resonator, simplifying the
equations modeling the array,
\begin{eqnarray}
  \label{eom2}
  \ddot{u}_{n} &+ &Q^{-1}\dot{u}_{n} +
  \bigl[1-H\cos(2\omega_{p}t)\bigr]u_{n} + u_{n}^{3} +
  \hat{\eta}u_{n}^{2}\dot{u}_n\nonumber\\
  &-&\frac{1}{2}D(u_{n+1}-2u_{n}+u_{n-1})=0.
\end{eqnarray}
The negative sign before the coupling coefficient $D$ models elastic
coupling between adjacent beams, which is stronger as the separation
between neighbors increases, thus acting to stiffen the
resonators. Here the dispersion curve has a positive slope, or a
positive group velocity.  The coupling mechanism in the experimental
setups in which ILMs have been observed is of this
kind~\cite{sato03_1,sato03_2,sato04,sato06,sato07,sato08}.

Because the quality factor $Q$ of a typical MEMS or NEMS resonator is
high we follow the practice~\cite{LCreview,LC,BCL} of using it to
define a small expansion parameter $Q^{-1}=\epsilon\hat{\gamma}$, with
$\epsilon\ll1$, and $\hat\gamma$ of order unity.  The driving
amplitude is then expressed as $H=\epsilon\hat{h}$, with $\hat{h}$ of
order unity, in anticipation of the fact that parametric oscillations
at half the driving frequency require a driving amplitude which is of
the same order as the linear damping rate~\cite{Landau}.

An experimental protocol for producing ILMs in an array of resonators
with a stiffening nonlinearity---albeit not the one we use below---is
to drive the array at the highest-frequency extended mode.  As the
resonators are collectively oscillating at this mode, the frequency is
raised further which results in an increase of the oscillation
amplitude up to a point in which the extended pattern breaks into
localized modes~\cite{sato03_1,sato06}.  With this in mind---and
concentrating on the case of elastic coupling where the
highest-frequency mode $\omega = \sqrt{1+2D}$ is the staggered mode,
in which adjacent resonators oscillate out of phase---we write the
displacement of the $n^{th}$ resonator as
\begin{equation}
  \label{u_expansion}
  \begin{split}
    u_{n} &= \epsilon^{1/2} \bigl[\hat{\psi}(\hat{X}_n,\hat{T})
    e^{i(\omega t - \pi n)} + c.c.\bigr]\\
    &\quad + \epsilon^{3/2}u_{n}^{(1)}(t,\hat{T},\hat{X}_n) + ...,
  \end{split}
\end{equation}
with slow temporal and spatial variables $\hat{T}=\epsilon t$ and
$\hat{X}_n = \epsilon^{1/2} n$, and $c.c.$ standing for the complex
conjugate expression.  We take the parametric drive frequency to be
close to twice $\omega$ by setting $\omega_{p} = \omega +\epsilon
\Omega/2$, introduce a continuous spatial variable $\hat{X}$ in place
of $\hat{X}_n$, and substitute the ansatz (\ref{u_expansion}) into the
equations of motion (\ref{eom2}) term by term. Up to order
$\epsilon^{3/2}$ we have
\begin{subequations}
\begin{align}
  &\ddot{u}_{n} = \epsilon^{1/2} \left[\left(-\omega^{2}\hat{\psi} +
    2i\omega\epsilon\frac{\partial\hat{\psi}}{\partial\hat{T}}\right)
  e^{i(\omega t - \pi n)}  + c.c.\right]\nonumber\\
  & \qquad + \epsilon^{3/2}\ddot{u}^{(1)}_{n}, \\
  &u_{n\pm1} = -\epsilon^{1/2} \Bigg[\left(\hat{\psi} \pm \epsilon^{1/2}
    \frac{\partial\hat{\psi}}{\partial\hat{X}} + \frac{\epsilon}{2}
    \frac{\partial^{2}\hat{\psi}}{\partial\hat{X}^{2}}\right)
  e^{i(\omega t-\pi n)}\nonumber\\
  & \hskip70pt  + c.c.\Bigg] + \epsilon^{3/2}u^{(1)}_{n\pm1},\\
  &\epsilon \hat{h}\cos(2\omega_{p}t)u_{n} =
  \epsilon^{3/2}\frac{\hat{h}}{2}\hat{\psi}^{*}e^{i\Omega\hat{T}}e^{i(\omega
  t+\pi n)}  + O(e^{i3\omega t})\nonumber\\
  & \hskip50pt + c.c., \\
  &\epsilon\hat{\gamma}\dot{u}_{n} =
  \epsilon^{3/2}\hat{\gamma}i\omega\hat{\psi}e^{i(\omega t - \pi n)} +
  c.c.,  \\
  &u_{n}^{3} =  \epsilon^{3/2}3|\hat{\psi}|^{2}\hat{\psi}e^{i(\omega
  t - \pi n)}  + O(e^{i3\omega t},e^{i3\pi n}) + c.c., \\
  &u_{n}^{2}\dot{u}_{n} =
  \epsilon^{3/2}i\omega|\hat{\psi}|^{2}\hat{\psi}e^{i(\omega t-\pi n)}
   + O(e^{i3\omega t},e^{i3\pi n}) + c.c.,
\end{align}
\end{subequations}
where $O(e^{i3\omega t},e^{i3\pi n})$ are fast oscillating terms with
temporal frequency $3\omega$ or spatial wavenumber $3\pi$.

At order $\epsilon^{1/2}$ the equations of motion~(\ref{eom2}) are
satisfied trivially. However, at order $\epsilon^{3/2}$, one must
apply a \emph{solvability condition}~\cite{reviewcross}, requiring all
terms proportional to $e^{i(\omega t-\pi n)}$ to vanish. It is this
condition that leads to a partial differential equation (PDE)
describing the slow dynamics of the amplitudes of the resonators,
\begin{equation}
  \label{sol_cond}
  2i\omega\frac{\partial \hat{\psi}}{\partial \hat{T}} + (3 +
  i\omega\hat{\eta})|\hat{\psi}|^{2}\hat{\psi} +
  \frac{1}{2}D\frac{\partial^{2}\hat{\psi}}{\partial \hat{X}^{2}} +
  i\hat{\gamma}\omega\hat{\psi} -
  \frac{\hat{h}}{2}\hat{\psi}^{*}e^{i\Omega \hat{T}} = 0.
\end{equation}
Note that while $e^{i(\omega t+\pi n)}=e^{i(\omega t-\pi n)}$, if we
were to consider an arbitrary mode of wave number $k$ instead of
$\pi$, the parametric term would have forced us to apply another
solvability condition, requiring terms proportional to $e^{i(\omega
  t+kn)}$ to vanish.  In that case an ansatz based on counter
propagating waves is considered as the $O(\epsilon^{1/2})$ solution
for $u_{n}$ and a system of two coupled amplitude equations emerges,
as shown in Ref.~\cite{BCL}.

By means of rescaling,
\begin{gather}\nonumber
  \hat{\psi} = \sqrt\frac{2 \omega\Omega}{3}\psi,\quad
  \hat{X}=\sqrt\frac{D}{2\omega\Omega} X,\quad
  \hat{T}=\frac{2}{\Omega} T,\\
  \hat{h}=2\omega\Omega h,\quad
  \hat{\gamma}=\Omega \gamma,\quad
  \hat{\eta}=\frac{3}{2\omega} \eta,
  \label{scaling}
\end{gather}
we transform Eq.~(\ref{sol_cond}) into a normalized form,
\begin{equation}
  \label{PDNLS}
  i\frac{\partial \psi}{\partial T} =
  - \frac{\partial^{2} \psi}{\partial X^{2}}
  - i\gamma\psi - (2 + i\eta)|\psi|^{2}\psi + h\psi^{*}e^{2iT}.
\end{equation}
We then perform one final transformation $\psi\rightarrow\psi e^{iT}$ and
arrive at an autonomous PDE, which is
the amplitude equation that we study in the remainder of this work,
\begin{equation}
  \label{amp_eq}
  i\frac{\partial \psi}{\partial T} =
  - \frac{\partial^{2} \psi}{\partial X^{2}}
  + (1 - i\gamma)\psi - (2 + i\eta)|\psi|^{2}\psi + h\psi^{*}.
\end{equation}
Equation (\ref{PDNLS}) with $\eta=0$ is called the
parametrically-driven damped nonlinear Schr\"odinger equation (PDNLS).
It models parametrically driven media in
hydrodynamics~\cite{zhang,wang97,wang98,miao} and
optics~\cite{longhi96,sanchez}, and was also used as an amplitude
equation to study localized structures in arrays of coupled
pendulums~\cite{denardo,chen,barashenkov00}. Recently, a pair of
linearly-coupled PDNLS equations was used to model coupled dual-core
wave guides~\cite{dror}. Eq.~(\ref{amp_eq}) has the form of a forced
complex Ginzburg-Landau equation~\cite{yochelis} but with specific
coefficients that are derived, via the scaling performed
in~(\ref{u_expansion}) and (\ref{scaling}), from the underlying
equations of motion~(\ref{eom2}).

We note that considering the equations of motion (\ref{eom}) (yet
still with a negative sign before $D$) instead of Eqs.~(\ref{eom2})
leads to the same equation~(\ref{sol_cond}) as above, but with
different coefficients (a factor of 2 multiplying $\hat{h}$ and
$\hat{\gamma}$, and a factor of 8 multiplying $\hat{\eta}$). Thus,
applying modified scaling~(\ref{scaling}) yields exactly the same
amplitude equation~(\ref{amp_eq}).

\section{SOLITONS IN THE PRESENCE OF NONLINEAR DAMPING}
\label{witheta}
\subsection{Continuation of the PDNLS solitons to $\eta>0$}
\label{continuation}

A remarkable feature the amplitude equation (\ref{amp_eq}) is that for
$\eta=0$ it has exact time-independent solitonic solutions, as shown
by Barashenkov \textit{et al.}~\cite{barashenkov91},
\begin{equation}
  \label{NLS_sol}
  \Psi_{\pm}(X) = A_{\pm}e^{-i\Theta{\pm}}
  \textrm{sech}\left[A_{\pm}\left(X - X_{0}\right)\right],
\end{equation}
where $X_{0}$ is an arbitrary position of the soliton, and
\begin{equation}
  \label{NLS_amp}
  A^{2}_{\pm} = 1\pm\sqrt{h^2 - \gamma^2},\
  \cos(2\Theta_{\pm}) =\pm\sqrt{1 - \frac{\gamma^{2}}{h^{2}}}.
\end{equation}
This pair of solitonic solutions exists for $\gamma<h$.  It was shown
in~\cite{barashenkov91} that the $\Psi_{-}$ soliton is unstable for
all values of $\gamma$ and $h$, while the $\Psi_{+}$ soliton is stable
in a certain parameter range.  A simple linear stability analysis
shows that the zero solution $\psi(X)=0$, which exists for all
parameter values, is stable only for $h<\sqrt{1+\gamma^{2}}$. This
inequality also determines an upper stability limit for localized
solutions of Eq.~(\ref{amp_eq}) that decay exponentially to zero on
either side.

The aim of this section is to show that the PDNLS solitons
$\Psi_{\pm}$ can be continued to nonzero nonlinear damping $\eta$.  A
similar calculation was performed by Barashenkov \emph{et
  al.}~\cite{barashenkov03} where they considered the addition of a
spectral filtering term $-ic \partial^{2}\psi/\partial X^{2}$ to the
PDNLS.  We do so by expanding the stationary solutions of the full
amplitude equation (\ref{amp_eq}) in powers of $\eta$, which is
assumed small, to get
\begin{equation}
  \label{expamsion}
  \psi(X) = (\phi_{0} + \eta\phi_{1} + \eta^{2}\phi_{2} + \ldots)
  e^{-i\Theta_{\pm}},
\end{equation}
so that $\phi_{0}=|\Psi_{\pm}|$.  Denoting $\phi_{n}=u_{n}+iv_{n}$, with
real $u_{n}$ and $v_{n}$, substituting $\psi$ into Eq.~(\ref{amp_eq}),
and comparing powers of $\eta$ yields equations of the form
\begin{equation}
  \label{lin_cont}
  L_{\pm}
  \begin{pmatrix}
    u_{n} \\
    v_{n} \\
  \end{pmatrix} =
  \begin{pmatrix}
      F_{n}(u_{0},v_{0},\ldots,u_{n-1},v_{n-1}) \\
      G_{n}(u_{0},v_{0},\ldots,u_{n-1},v_{n-1}) \\
  \end{pmatrix},
\end{equation}
where
\begin{equation}
  L_{\pm} =
  \begin{pmatrix}
    -\partial^{2}_{X} - 6u_{0}^{2} + A_{\pm}^{2} & 2\gamma \\
    0 & -\partial^{2}_{X} - 2u_{0}^{2} + A_{\mp}^{2} \\
  \end{pmatrix}.
\end{equation}

One can use Eq.~(\ref{lin_cont}) iteratively to find the
$n^{th}$-order correction $\phi_n$, given all lower-order ones,
provided that the right-hand side is orthogonal to the null subspace,
or kernel, of the adjoint operator $L_{\pm}^{\dagger}$, if such a
subspace exists. Indeed, in the relevant parameter range,
$\gamma<h<\sqrt{1+\gamma^{2}}$, the adjoint operator
$L_{\pm}^{\dagger}$ has one zero eigenvalue, and the corresponding
eigenvector consists of only odd functions of
$X-X_{0}$~\cite{barashenkov03,zemlyanaya}.  On the other hand, one can
verify that the functions $F_{n}$ and $G_{n}$, which originate from
the nonlinear terms in Eq.~(\ref{amp_eq}), include only natural powers
of $u_{0},v_{0},\ldots,u_{n-1},v_{n-1}$. Therefore, and because
$u_0=|\Psi_{\pm}|$, $v_0=0$, and $L_\pm$ is parity preserving, the
right hand side of Eq.~(\ref{lin_cont}) consists of only even
functions of $X-X_{0}$, for any $n$.  This suggests that it is
possible to continue the PDNLS solitons $\Psi_{\pm}$ to solve
Eq.~(\ref{amp_eq}) up to any order in $\eta$.  This can also be done
in practice, by calculating $F_{n}$ and $G_{n}$ symbolically,
expressing $L_{\pm}$ as a discrete matrix, and inverting it to find
$u_{n}$ and $v_{n}$ in each iteration, although we do not follow this
procedure here.

\subsection{Approximate analytical solitons with $\eta>0$}
\label{solution}

Motivated by the arguments above, we wish to construct an approximate
analytical expression for the localized solution of the full amplitude
equation (\ref{amp_eq}), implementing the method of Barashenkov
\emph{et al.}~\cite{barashenkov03}. To this end, we consider a
function of the same form as $\Psi_{\pm}$,
\begin{equation}
  \label{aprx_sol_time_dependant}
  \psi(X,T) = a(T) e^{-i\theta(T)}
  \textrm{sech}\left[a(T) \left(X-X_0\right)\right],
\end{equation}
except that $a$ and $\theta$ are now time-dependent. We multiply
Eq.~(\ref{amp_eq}) by $\psi^{*}$, subtract the complex conjugate of
the resulting equation and get
\begin{eqnarray}
  i\frac{\partial|\psi|^{2}}{\partial T} &= &-\frac{\partial}{\partial
    X}{\left(\frac{\partial \psi}{\partial X}\psi^{*} -
      \psi\frac{\partial \psi^{*}}{\partial X}\right)} +
  h[(\psi^{*})^{2} - \psi^{2}]\nonumber\\
  &- &2i\gamma|\psi|^{2} - 2i\eta|\psi|^{4}.
\end{eqnarray}
By substituting $\psi=|\psi|e^{-i\chi}$, integrating over $X'=X-X_0$,
and assuming that $\psi\rightarrow0$ and $\partial\psi/\partial
X\rightarrow0$ as $|X|\rightarrow\infty$, we obtain a
spatially-independent integral equation
\begin{equation}
  \label{no_x_eq}
  \frac{d}{dT}\int|\psi|^{2}dX' = 2\int|\psi|^{2}[h\sin(2\chi) -
  \gamma]dX' - 2\eta\int|\psi|^{4}dX'.
\end{equation}
Substituting the ansatz~(\ref{aprx_sol_time_dependant}) into
Eq.~(\ref{no_x_eq}), we obtain the time evolution equation for $a$
\begin{equation}
  \label{a_T}
  \frac{d a}{dT}=2a(h\sin(2\theta) - \gamma - \tilde{\eta}a^{2}),
\end{equation}
where $\tilde{\eta}=2\eta/3$. The time evolution equation for $\theta$
is derived in a similar way by multiplying Eq.~(\ref{amp_eq}) by
$\psi^{*}$, adding the complex conjugate of the resulting equation,
substituting the ansatz~(\ref{aprx_sol_time_dependant}), and
integrating over space to yield
\begin{equation}
  \label{theta_T}
  \frac{d\theta}{dT} = h\cos(2\theta) + 1 - a^{2}.
\end{equation}

Equations (\ref{a_T}) and (\ref{theta_T}) have the same form as the
equations obtained in~\cite{barashenkov03}, whose fixed points are
\begin{equation}
  \label{aprx_amp}
  a^{2}_{\pm} = \frac{1 - \gamma\tilde{\eta}
    \pm\sqrt{h^{2}(1 + \tilde{\eta}^{2}) - 
      (\gamma + \tilde\eta)^2}}{1 + \tilde{\eta}^{2}},
\end{equation}
which has to be positive, and
\begin{eqnarray}
  \label{aprx_phase}\nonumber
  h\cos(2\theta_{\pm}) &= &a_{\pm}^{2} - 1,\\
  h\sin(2\theta_{\pm}) &= &\gamma + \tilde\eta a_{\pm}^{2}.
\end{eqnarray}
A linear analysis of  these stationary
points shows that $(a_{+},\theta_{+})$ and $(a_{-},\theta_{-})$ are a
stable node and a saddle, respectively~\cite{barashenkov03}. The
saddle-node bifurcation point of these solutions occurs at
\begin{equation}
  \label{sn_g}
  h_{sn}(\tilde\eta) = \frac{\gamma + \tilde{\eta}}{\sqrt{1 +
  \tilde{\eta}^{2}}}, 
  \quad \textrm{where} \quad \tilde{\eta}=\frac{2}{3} \eta,
\end{equation}
as long as $\gamma\tilde\eta<1$.  This is the approximate minimal
driving strength required to support a localized structure in the
array, in the presence of linear and nonlinear dissipation.

The approximate stable localized solution of the amplitude
equation~(\ref{amp_eq}) is therefore given by
\begin{equation}
  \label{aprx_sol}
  \pap(X) = a_{+}e^{-i\theta_{+}}\textrm{sech}(a_{+}(X - X_{0})).
\end{equation}
Substituting this expression into Eq.~(\ref{u_expansion}) yields an
approximate expression for the displacements of the actual resonators
in the array,
\begin{eqnarray}
  \label{u_sol}
  u_{n}(t) &\simeq &2\sqrt{\frac{2\epsilon\omega\Omega}{3}}a_{+}
  \textrm{sech}\left[a_{+}\left(\sqrt{\frac{2\epsilon\omega\Omega}{D}}n
      - X_{0}\right)\right]\nonumber\\
  & \times &\cos\left(\omega_{p}t - \pi n - \theta_{+}\right).
\end{eqnarray}

\subsection{Numerical solutions for solitons with $\eta>0$}
\label{exactsolitons}

To obtain accurate solutions we solve the amplitude equation, as well
as the underlying discrete equations of motion, numerically. The
equations of motion~(\ref{eom2}) are initiated with the approximate
expression~(\ref{u_sol}) at a value of $h$ just above the saddle node
$h_{sn}$~(\ref{sn_g}). We then perform a quasistatic upward sweep of
$h$, raising $h$ in small increments and waiting for transients to
decay at each step.  To obtain the stationary solution of the
amplitude equation we set $\partial\psi/\partial T=0$ in
Eq.~(\ref{amp_eq}) and solve it numerically as a boundary value
problem over an interval of length $L$, with boundary conditions
$\psi(X=0)=\psi(X=L)=0$. We use the approximate expression $\pap(X)$
[Eq.~(\ref{aprx_sol})] as an initial guess.

\begin{figure}
\includegraphics[width=0.85\columnwidth]{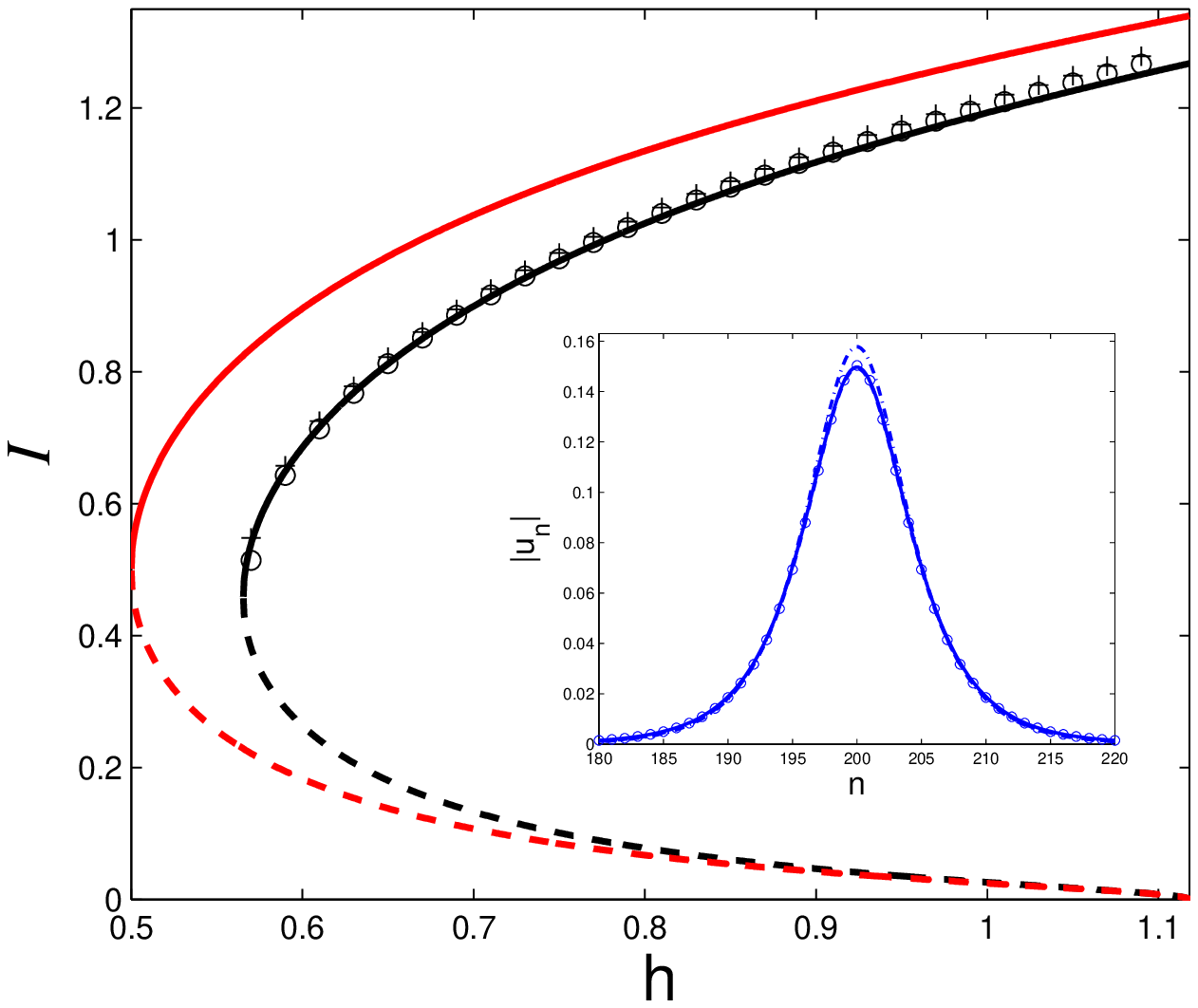}%
\caption{\label{sweep_inset}(Color online) The integral measure
  $I\{f(X)\}$ as a function of $h$.  Red outer solid and dashed lines
  represent the exact analytical solutions of the PDNLS equation
  without nonlinear damping $A_{+}\cos^{2}\Theta_{+}$ and
  $A_{-}\cos^{2}\Theta_{-}$, respectively. Black inner solid and
  dashed lines represent the approximate analytical solutions with
  nonlinear damping $a_{+}\cos^{2}\theta_{+}$ and
  $a_{-}\cos^{2}\theta_{-}$, respectively. These analytical lines end
  at the upper stability boundary $h=\sqrt{1+\gamma^{2}}$. The points
  designated by crosses and circles ($+$ and $\circ$) are taken,
  respectively, from the stationary numerical solution of the
  amplitude equation~(\ref{amp_eq}), and from the numerical solution
  of the equations of motion~(\ref{eom2}), as elaborated in the
  text. The inset shows the absolute value of the profile of the
  solution for $h=0.87$ where $\circ$s are results of the numerical
  solution of the equations of motion (\ref{eom2}). The solid line
  shows the real part of the numerical solution $\psi$ of the
  amplitude equation (\ref{amp_eq}), scaled by a factor of
  $2\sqrt{2\epsilon\omega\Omega/3}$. The scaled analytical
  approximation~(\ref{aprx_sol}) is indistinguishable from the solid
  line in this plot. The dot dashed line shows the scaled analytical
  solution~(\ref{NLS_sol}) in the absence of nonlinear damping. The
  parameters are $\gamma=0.5, D=0.25, \epsilon=0.01,
  \omega_{p}=1.002\omega, \eta=0.1$, and $N=399$.}
\end{figure}

In Fig.~\ref{sweep_inset} we compare the integral measure
\begin{equation}
  \label{measure}
  I\{f(X)\} = \frac12 \int_{-\infty}^{\infty} f^{2}(X) dX
\end{equation}
for the different localized solutions, where for the analytical
solutions $f(X) = a \textrm{sech}(aX) \cos(\theta)$ and
$I=a\cos^{2}(\theta)$, whereas for the stationary numerical solution
$\psi(X)$ of the amplitude equation~(\ref{amp_eq}) $f(X) = {\rm
  Re}\psi(X)$, and for the numerical steady-state solution of the
equations of motion~(\ref{eom2}) $f(X)$ is the appropriately scaled
magnitude $|u_n|$ of the $n^{th}$ resonator measured at times
$t_{m}=2\pi m/\omega_{p}$. Fig.~\ref{sweep_inset} shows good agreement
between the numerical solution of the equations of motion
(\ref{eom2}), and the approximate and numerical solutions of the
amplitude equation (\ref{amp_eq}).

\begin{figure}
\includegraphics[width=0.95\columnwidth]{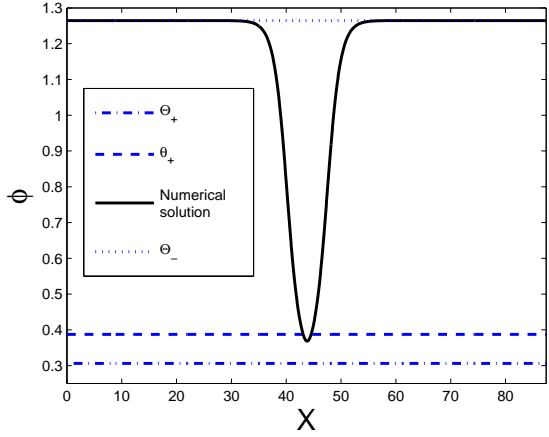}%
\caption{\label{phase}(Color online) The phase of the numerical
  solution of the amplitude equation (\ref{amp_eq}), and the
  analytical exact and approximate phases, as indicated in the legend.
  Near the peak of the soliton, the phase of the numerical solution is
  close to $\theta_{+}$ and it asymptotes to $\Theta_{-}$ as the
  soliton's amplitude decays to zero. Parameters are the same as in
  the inset of Fig.~\ref{sweep_inset}.}
\end{figure}

Fig.~\ref{phase} shows the phase of the numerical solution of the
amplitude equation~(\ref{amp_eq}), calculated as
\begin{equation}
  \label{eq:numericalphase}
  \phi = -\arctan \left(\frac{\textrm{Im}\psi}{\textrm{Re}\psi}\right).
\end{equation}
One can see that while the approximate phase $\theta_{+}$ provides a
good estimate for the actual phase of the solution near the peak of
the soliton, the phase asymptotes to $\Theta_{-}=\pi/2-\Theta_{+}$ as
the soliton's amplitude decays to zero. This is a surprising result
since it might have been expected that the phase of a numerical
continuation of the $\Psi_{+}$ solution would tend back to
$\Theta_{+}$ as the amplitude drops to zero, eliminating the nonlinear
damping. However, a similar situation was observed in a bound state of
two $\Psi_{+}$ solitons in the PDNLS equation~\cite{barashenkov99}.

\section{LINEAR STABILITY ANALYSIS OF SOLITONS} 
\label{stability}

\begin{figure}
\includegraphics[width=\columnwidth]{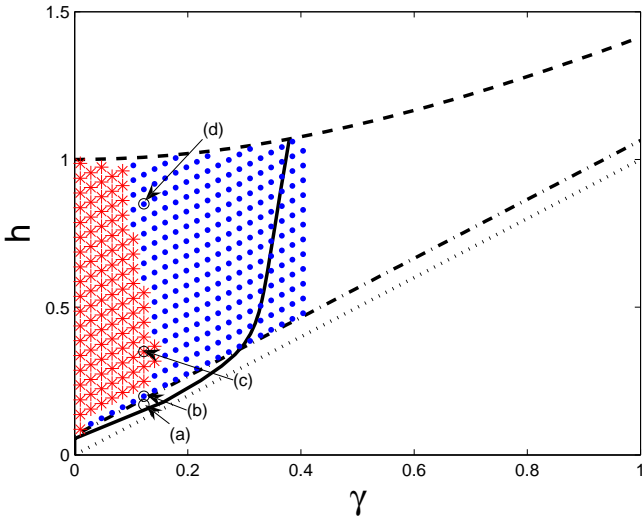}%
\caption{(Color online) Stability diagram for localized solutions of
  the amplitude equation~(\ref{amp_eq}) in the $h$ vs.~$\gamma$ plane.
  The dotted line is the lower existence boundary for $\eta=0$, namely
  $h=\gamma$. The dashed-dotted line is the approximate low boundary
  for $\eta=0.1$, given by Eq.~(\ref{sn_g}). Above the solid line the
  $\Psi_{+}$ solution of the PDNLS equation with $\eta=0$ is unstable
  with respect to a Hopf bifurcation~\cite{barashenkov91}.  The dashed
  line is the line $h=\sqrt{1+\gamma^{2}}$ above which the zero
  solution is unstable. Red $\ast$s are points for which the matrix
  $J^{-1}H$ has a pair of complex conjugate eigenvalues with a
  positive real part for $\eta=0.1$, hence the soliton solution
  $\psi(X)$ is unstable. Blue dots represent points for which the
  solution $\psi(X)$ is stable according to the linear analysis. The
  four black circles labeled (a)-(d) indicate parameter values
  corresponding to the numerical simulations of the equations of
  motion (\ref{eom2}) shown in
  Figs.~\ref{simulations}(a)-\ref{simulations}(d).\label{stability_diagram}}
\end{figure}

\begin{figure}
    \begin{center}
    \subfigure[]{
    \includegraphics[width=0.48\columnwidth]{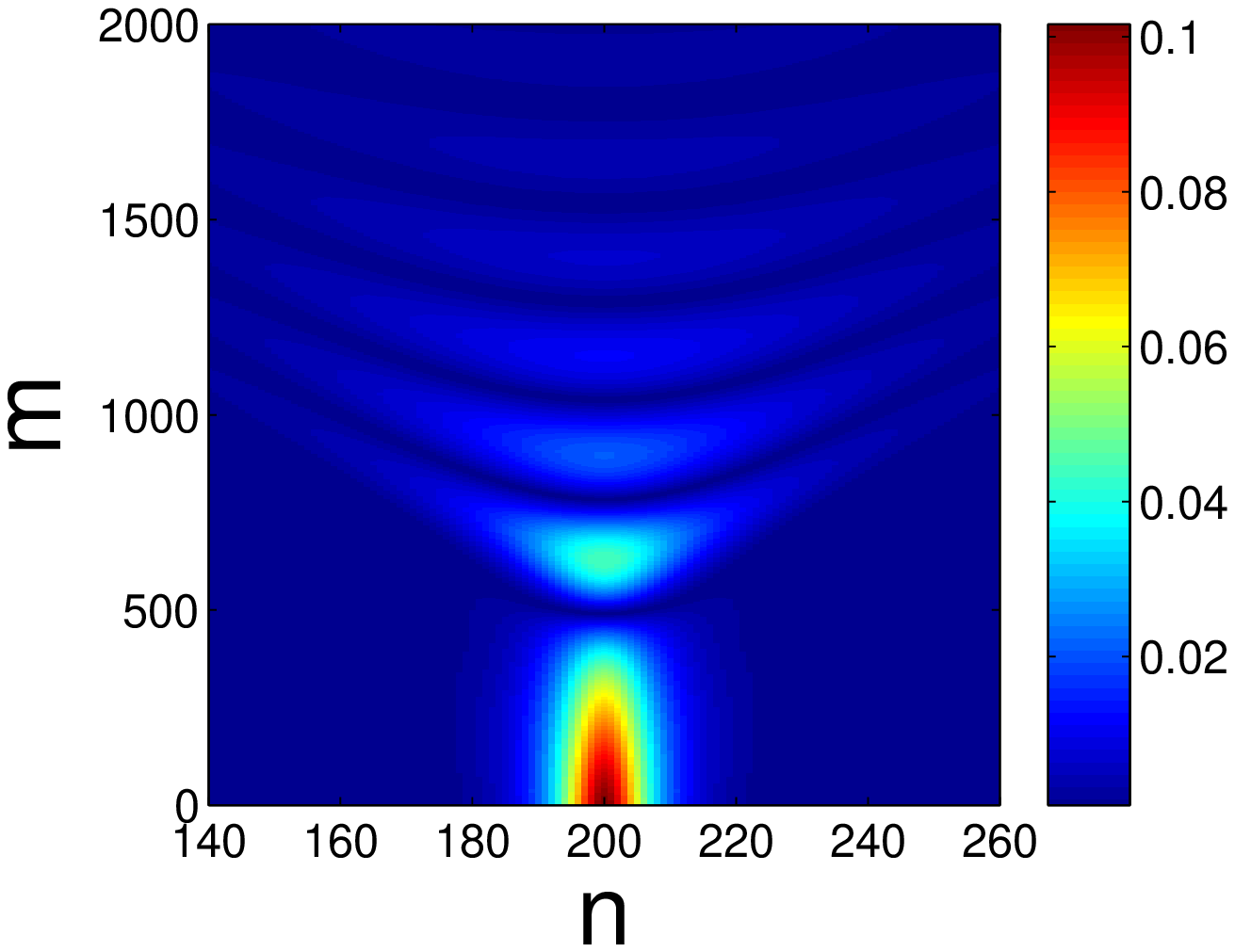}}
    \subfigure[]{
    \includegraphics[width=0.48\columnwidth]{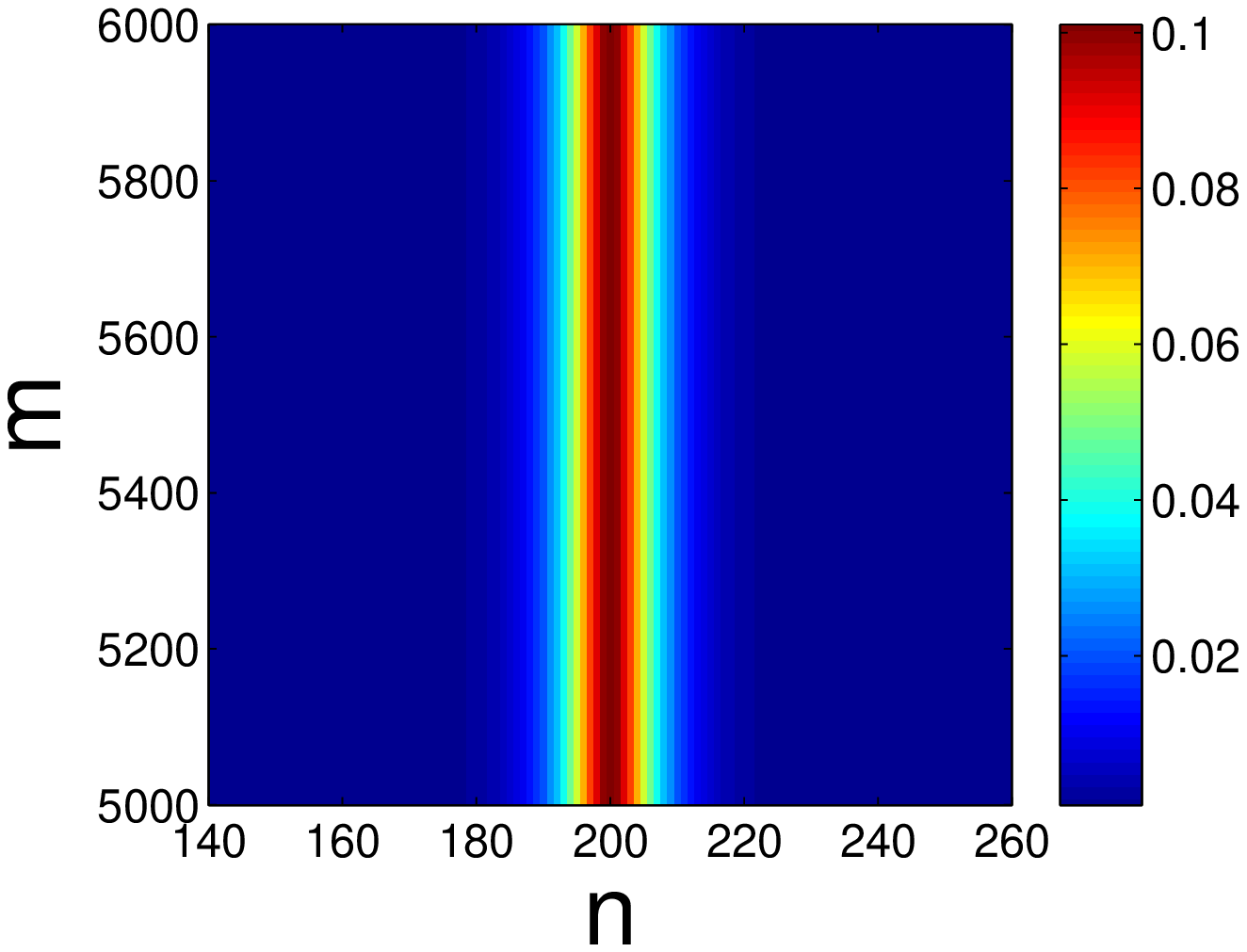}}
    \subfigure[]{
    \includegraphics[width=0.48\columnwidth]{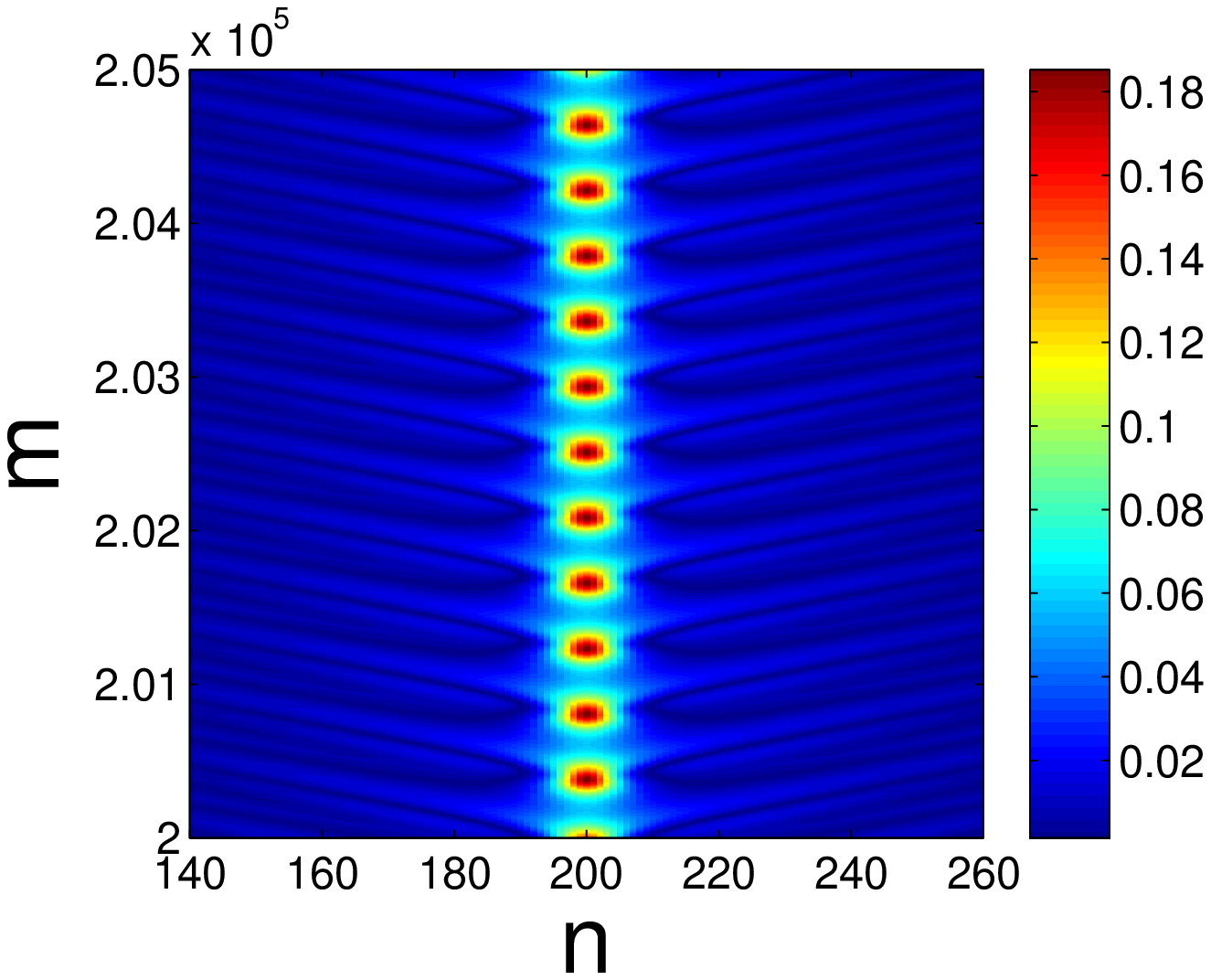}}
    \subfigure[]{
    \includegraphics[width=0.48\columnwidth]{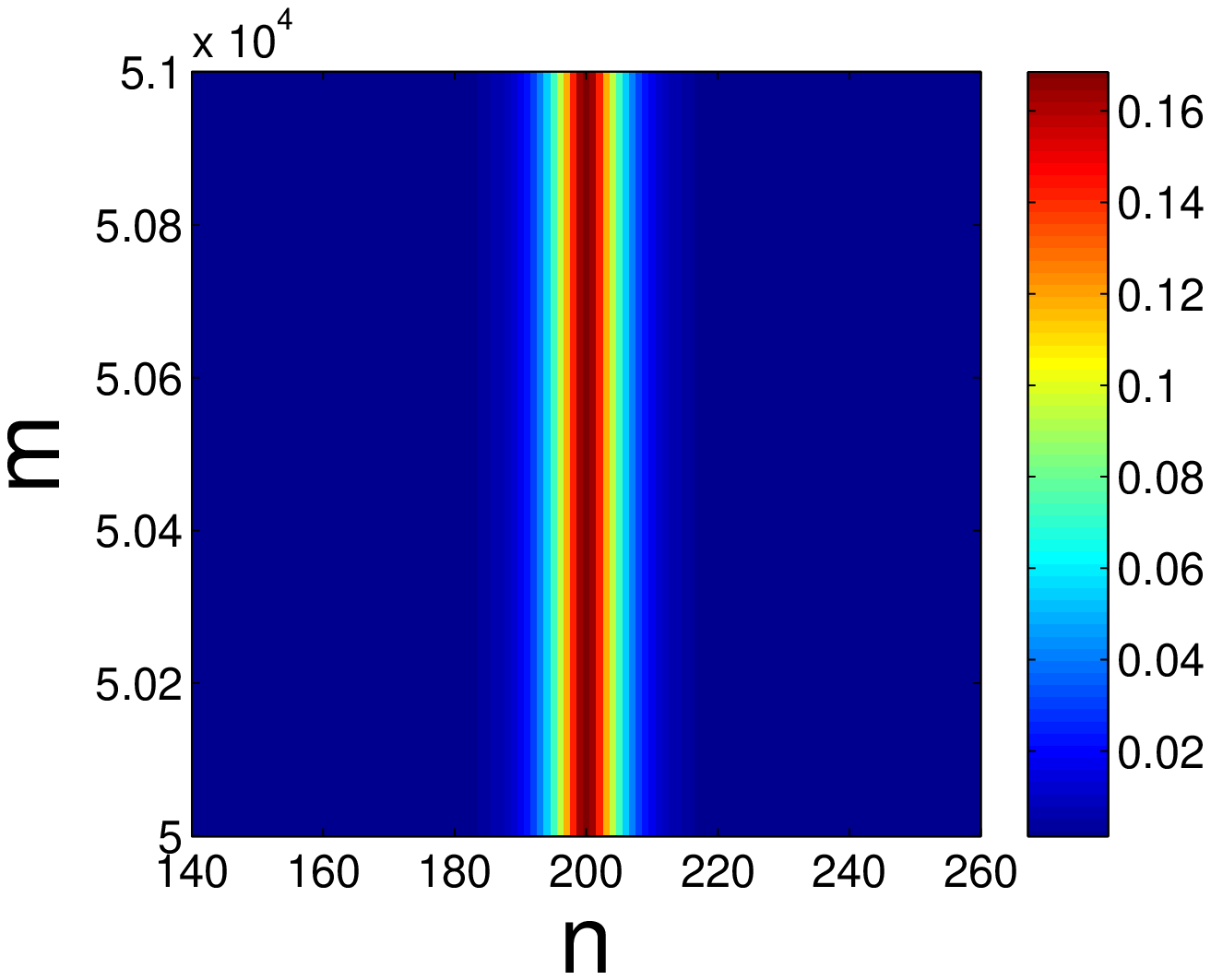}}
    \subfigure[]{
    \includegraphics[width=0.48\columnwidth]{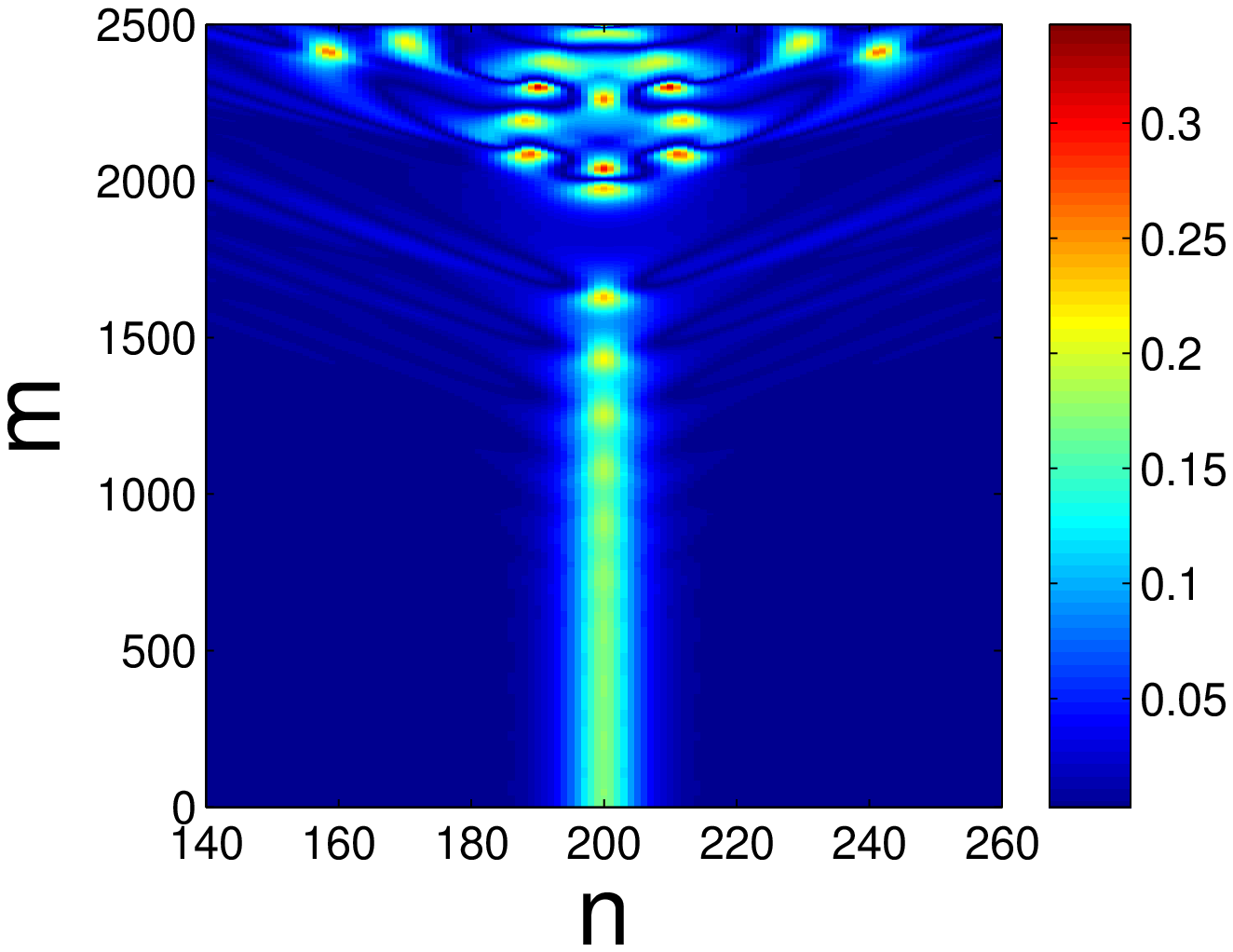}}
    \end{center}
    \caption{\label{simulations}(Color online) Results of the
      numerical solution of the equations of motion (\ref{eom2}) for
      $\eta=0.1$, $\gamma=0.1225$, and values of $h$ labeled as
      (a)-(d) in Fig.~\ref{stability_diagram}.  The solutions are
      plotted at times $t_{m}=2\pi m/\omega_{p}$, with integer values
      $m$ shown on the vertical axis. (a) $h=0.17$ is below the
      approximate low boundary~(\ref{sn_g}) but above the low boundary
      for $\eta=0$. One sees that the localized structure decays to
      zero. (b) $h=0.1988$ is above the approximate low boundary,
      where linear stability analysis predicts that the soliton is
      stable. (c) For $h=0.35$ the stationary soliton is unstable, and
      an oscillating localized solution is formed instead. In (d)
      $h=0.85$ and the soliton is stable again. The stability is due
      to nonlinear damping, without which the soliton is unstable as
      demonstrated in (e) where $h=0.85$ and $\eta=0$. All unspecified
      parameters are the same as in Fig.~\ref{sweep_inset}.}
\end{figure}

Having identified an upper stability boundary $h=\sqrt{1+\gamma^{2}}$
and an approximate lower existence boundary, given by
Eq.~(\ref{sn_g}), we turn to examine the stability of the localized
solution within these boundaries. For this purpose, we substitute into
Eq.~(\ref{amp_eq}) $\psi(X,T)=\psi(X)+\delta\psi(X,T)$, where
$\psi(X)$ could be any steady-state solution of the equation---in this
case the stationary localized solution, which is obtained
numerically---and $\delta\psi(X,T)$ is a small perturbation. We
linearize in $\delta\psi(X,T)$, substituting $\psi=R+iI$ and
$\delta\psi=U+iV$ with real $R,I,U,$ and $V$, and obtain the equation
\begin{equation}
  \label{H_matrix}
  J \frac{\partial}{\partial T}
  \begin{pmatrix}
    U \\
    V \\
  \end{pmatrix} = H
  \begin{pmatrix}
    U \\
    V \\
  \end{pmatrix},
\end{equation}
where
\begin{equation}
  \label{H_matrix_def}
  J =
  \begin{pmatrix}
    0 & -1 \\
    1 & 0 \\
  \end{pmatrix},\qquad
  H =
  \begin{pmatrix}
    H_{11} & H_{12}\\
    H_{21} & H_{22}\\
  \end{pmatrix},
\end{equation}
\begin{eqnarray}\label{H}
  H_{11} &=& -\partial^{2}_{X} - 6R^{2} - 2I^{2} + 1 + h + 2\eta RI,\nonumber\\
  H_{12} &=& \eta(R^{2} + 3I^{2}) - 4RI + \gamma,\nonumber\\
  H_{21} &=& -\eta(3R^{2} + I^{2}) - 4RI - \gamma,\nonumber\\
  H_{22} &=& -\partial^{2}_{X} - 6I^{2} - 2R^{2} + 1 - h - 2\eta RI.
\end{eqnarray}
By expressing the small perturbations as $U(X,T)={\rm Re}[u(X)
e^{\lambda T}]$ and $V(X,T)={\rm Re}[v(X) e^{\lambda T}]$, where
$\lambda,u,$ and $v$ are complex, we arrive at the eigenvalue problem
\begin{equation}
  \label{H_matrix_complex}
  \lambda J
  \begin{pmatrix}
    u \\
    v \\
  \end{pmatrix}
  = H
  \begin{pmatrix}
    u \\
    v \\
  \end{pmatrix}.
\end{equation}
When $\psi(X)$ is obtained numerically, the eigenvalues of the matrix
$J^{-1}H$, describing the growth of perturbations $\delta\psi(X,T)$,
are found by performing a spatial discretization of
Eq.~(\ref{H_matrix_complex}) and diagonalizing $J^{-1}H$ numerically
(see~\cite{barashenkov96} for details).  The stability diagram of both
the analytical solution $\Psi_{+}$ for $\eta=0$~\cite{barashenkov91}
and the numerical solution $\psi(X)$ for $\eta=0.1$ are displayed in
Fig.~\ref{stability_diagram}.  These results are verified at a few
points by numerical integration of the equations of
motion~(\ref{eom2}), as shown in Fig.~\ref{simulations}.

Fig.~\ref{stability_diagram} highlights the effects of nonlinear
damping on localized solutions. The first effect is to raise the lower
existence boundary. This is explained by the fact that the additional
energy lost through nonlinear damping has to be compensated by an
increase in the strength of the parametric drive, as predicted by the
approximate expression~(\ref{sn_g}).  The second effect is that
nonlinear damping increases the area in the $(h,\gamma)$ parameter
space where solitons are stable (blue dots). In particular, the shape
of the unstable region for $\eta>0$ (red $\ast$s) becomes
qualitatively different. There are values of $\gamma$ for which an
increase in the drive amplitude $h$ initially induces an instability
of the soliton, while upon further increase of $h$ the soliton regains
its stability. This can be explained by noting that the amplitude of
the soliton---given approximately by Eq.~(\ref{aprx_amp})---increases
as $h$ becomes larger, thereby enhancing the effect of nonlinear
damping. This increase of damping exerts a similar stabilizing effect
as that of increasing $\gamma$ in the absence of nonlinear damping.

The effect of regained stability with the increase of $h$ does not
occur in the absence of nonlinear damping, as the $\Psi_{+}$ soliton
is unstable for all parameter values above the solid black line in
Fig.~\ref{stability_diagram}~\cite{barashenkov91}. Different solutions
of the PDNLS equation above this instability threshold were found by
Bondila \textit{et al.}~\cite{bondila} to be localized solutions that
oscillate in time with different periods and chaotic solutions, in
addition to the zero solution. As we increase the nonlinear damping
coefficient $\eta$, the region in $(h,\gamma)$ space, in which the
single soliton becomes unstable against these alternative solutions,
shrinks in size.

\section{DYNAMICAL FORMATION OF SOLITONS}
\label{flat}
\subsection{Self-trapping of solitons}

\begin{figure}
    \centering
    \subfigure{
    \includegraphics[width=0.48\columnwidth]{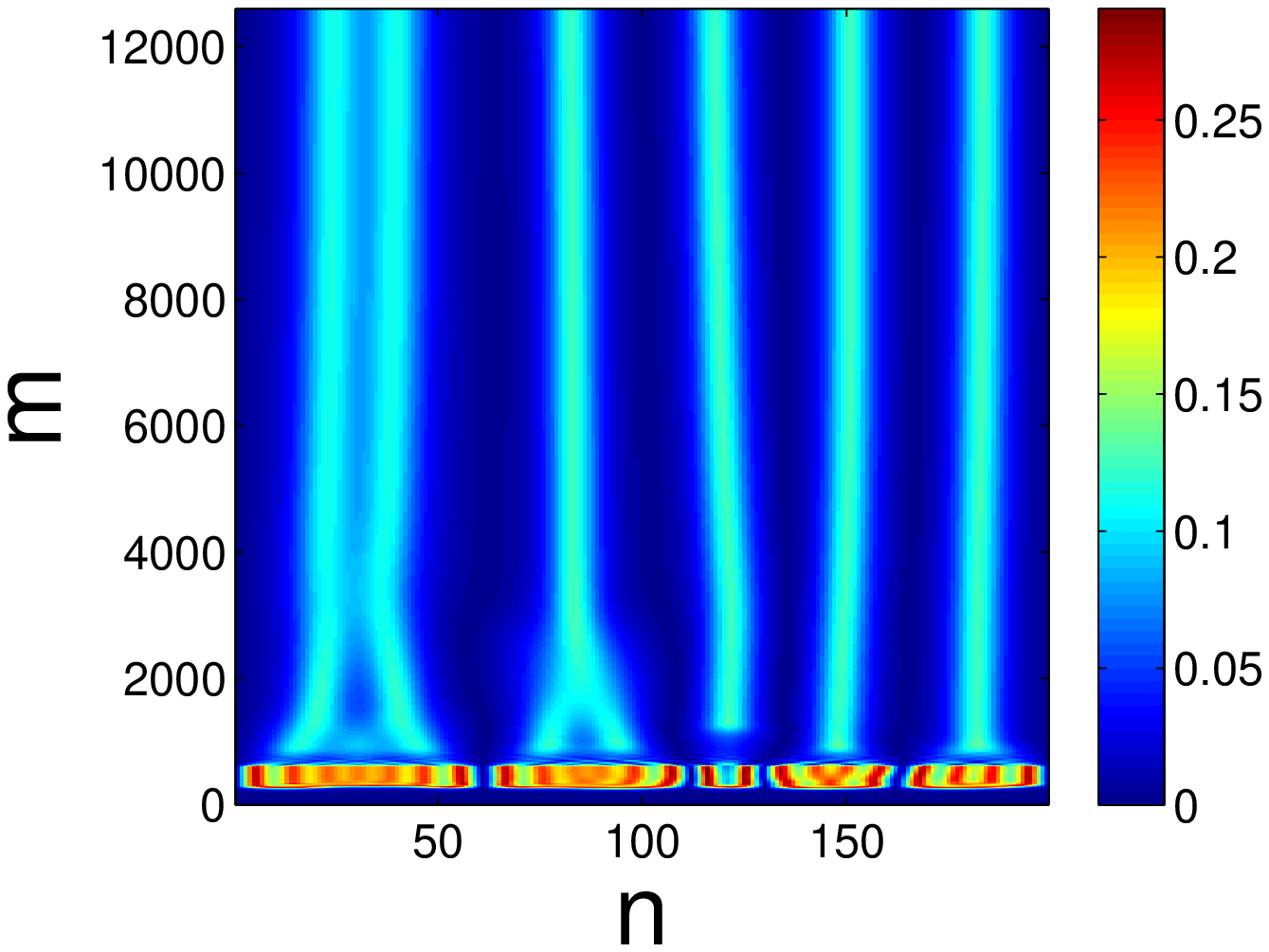}}
    \subfigure{
    \includegraphics[width=0.48\columnwidth]{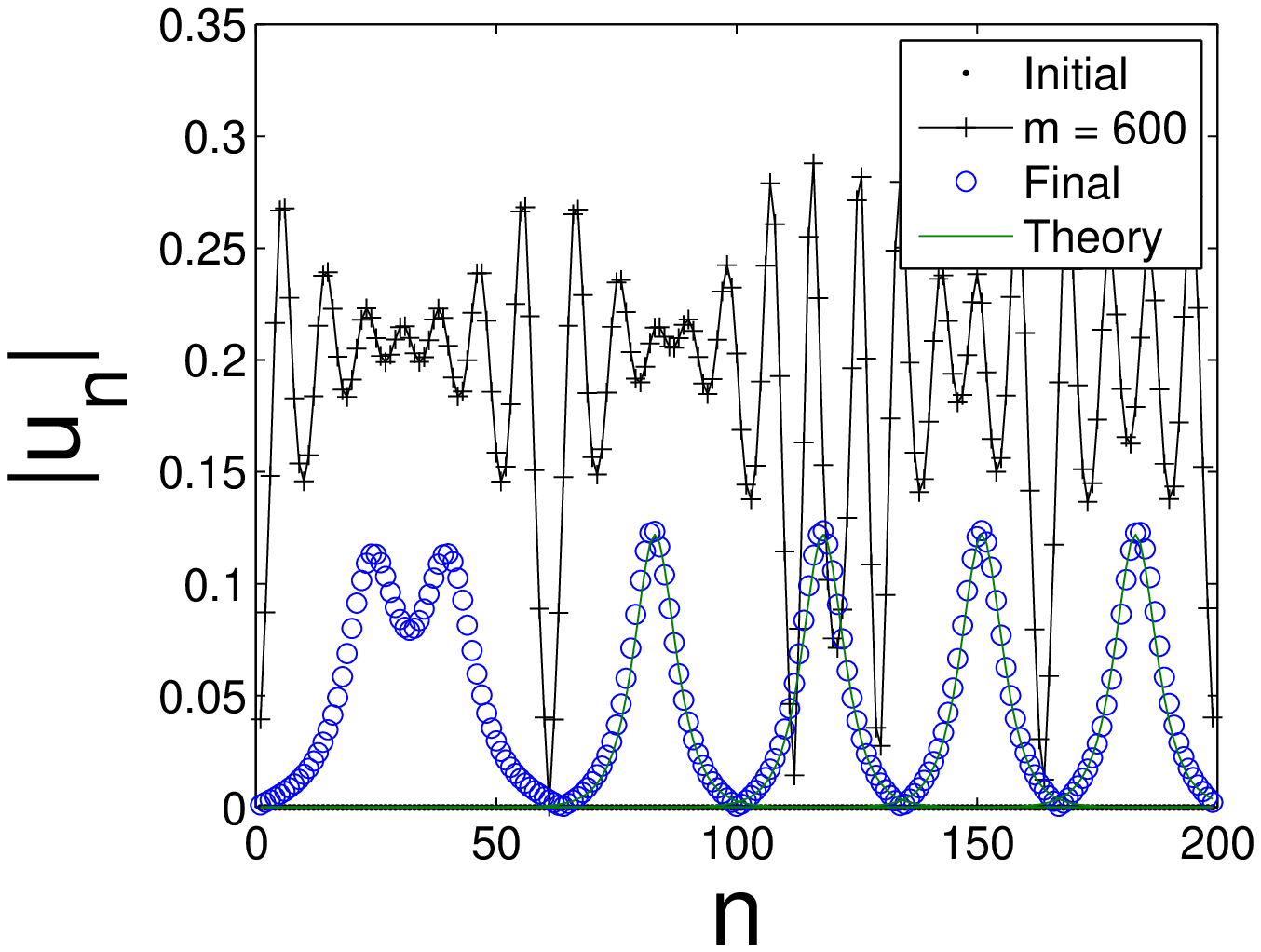}}
    \subfigure{
    \includegraphics[width=0.48\columnwidth]{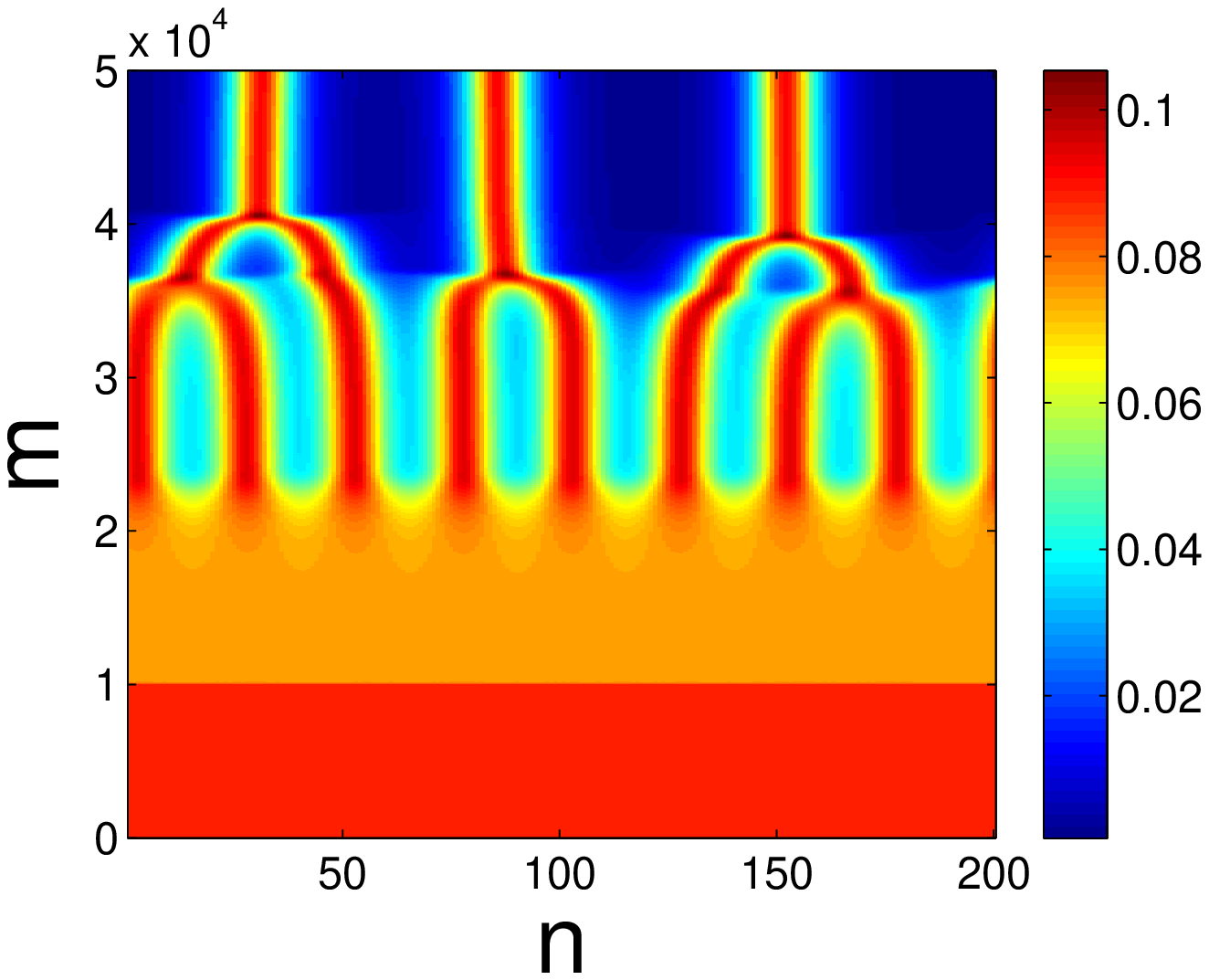}}
    \subfigure{
    \includegraphics[width=0.48\columnwidth]{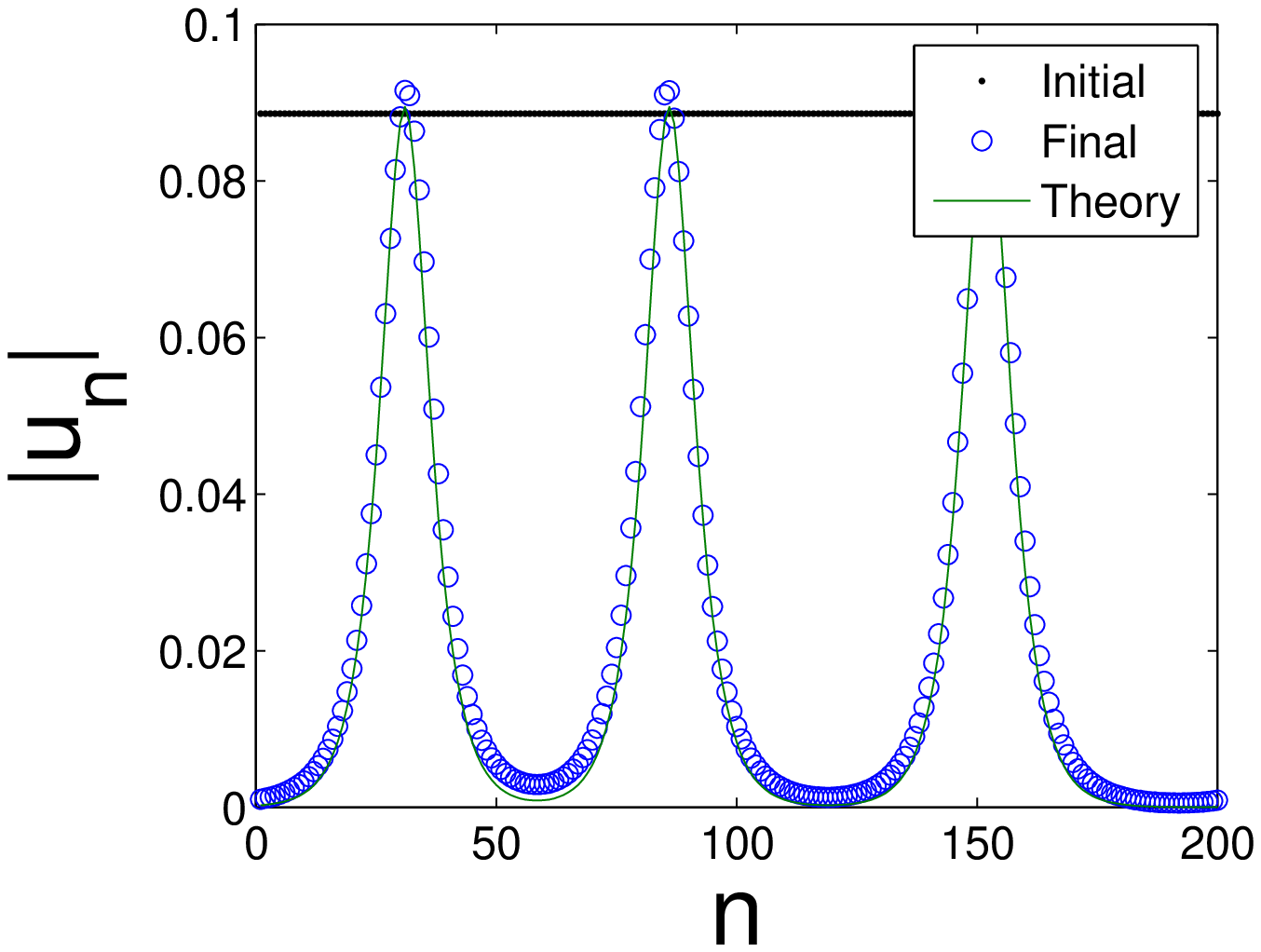}}
  \caption{\label{ilm creation}(Color online) Numerical simulation of
    the coupled equations of motion~(\ref{eom2}) showing the dynamical
    creation of solitons. Linear damping is set to $\gamma=1$ and
    nonlinear damping to $\eta=0.3$. All other parameters are the same
    as in Fig.~\ref{sweep_inset}.  Plotted are the absolute values of
    the displacements of the resonators, which alternate between
    positive and negative values.  Left panels show the complete time
    evolution, with $m$ counting the number of drive periods. Right
    panels show the initial (black dots) and final (blue circles)
    states along with the analytical form of the solitons (green solid
    line), using only their central positions $X_0$ as fitting
    parameters. \emph{Top panels}: A simulation of 199 resonators with
    fixed boundary conditions is initiated with random noise and a
    drive amplitude of $h=5$, which is above the upper stability
    limit, $h=\sqrt{1+\gamma^{2}}=\sqrt2$, for both the zero-state and
    the solitons.  At time $m=600$ drive periods, after some non-zero
    transient (black $+$s in the right panel) has developed, the drive
    amplitude is lowered to $h=1.35<\sqrt2$, yielding stable solitons.
    \emph{Bottom panels}: A simulation of 200 resonators with periodic
    boundary conditions is initiated with the uniform non-zero
    solution and a drive amplitude of $h=1.3$, which is above the
    stability threshold~(\ref{eq:htheshold}), $h_{th}\simeq1.26$, for
    this state.  After $m=10000$ drive periods during which the uniform
    state remains stable, the drive amplitude is lowered to
    $h=1.2<h_{th}$, yielding stable solitons.  }
\end{figure}

It is not obvious how dynamically to form solitons starting with a
motionless array of resonators, as one needs to take the system
sufficiently far from the basin of attraction of the zero solution
$\psi(X)=0$, which is also stable whenever solitons are stable. The
most direct procedure for avoiding the zero solution, starting from
weak random noise, is to drive the system with
$h>\sqrt{1+\gamma^{2}}$, so neither the zero solution nor the soliton
solutions are stable. As a consequence, a non-zero pattern develops.
Stable solitons can then be formed by lowering the drive amplitude
to a value $h<\sqrt{1+\gamma^{2}}$ for which the zero solution and the
soliton solutions are both stable, if the non-zero pattern that was
obtained is outside the basin of attraction of the zero solution. 

This simple procedure---which could be implemented experimentally in
a straightforward manner---is demonstrated in the top panels of
Fig.~\ref{ilm creation}, showing a numerical simulation of the
equations of motion~(\ref{eom2}) with fixed boundary conditions, using
$N=199$ resonators. One can see that the initial transient that forms
becomes unstable upon lowering the drive amplitude, giving rise to the
formation of a number of solitons. Note that before reaching steady
state a pair of solitons merges into one, and another pair attracts
and forms a bound state. Both of these effects are studied below. The
emerging isolated solitons agree well with the approximate analytical
form~(\ref{u_sol}), determined earlier, with only their central
positions $X_0$ used as fitting parameters.

\subsection{Modulational instability of uniform states}

A more controlled procedure for generating solitons would be to
initiate the array in a particular non-zero state and drive it outside
its known stability boundaries. This has been considered in the past
in systems without nonlinear damping, using the non-zero uniform
solution of the PDNLS~\cite{barashenkov03,wang,thakur,maniadis}.
However, it is known for systems with $\eta=0$ that the uniform
solution is always unstable against weak modulations and so may be
difficult to access dynamically. We wish to examine here whether the
non-zero uniform solution may be stabilized with the help of nonlinear
damping ($\eta\neq0$), thereby making it accessible dynamically and
possibly opening an additional experimental route to the formation of
solitons.

Indeed, the amplitude equation~(\ref{amp_eq}) admits a pair of
non-zero spatially-uniform solutions of the form
\begin{equation}\label{uniform}
    \bar{\psi}_{\pm}=\bar{a}_{\pm}e^{-i\bar{\theta}_{\pm}}.
\end{equation}
Substituted into the perturbative expansion~(\ref{u_expansion}), this
yields an oscillation of the array in its staggered mode with
wavenumber $\pi$, about which we initially expanded our solution. If
we impose fixed boundary conditions the staggered mode will be
modified near the boundaries to accommodate these conditions, but would
otherwise remain unchanged in the bulk of the system. 

Letting $\bar\eta=\eta/2$ and substituting the uniform solution
(\ref{uniform}) into the amplitude equation~(\ref{amp_eq}) yields
\begin{equation}
  \label{eq:AMPuniform}
  2\bar{a}^{2}_{\pm} = \frac{1 - \gamma \bar\eta \pm \sqrt{h^{2}(1 +
   \bar\eta^{2}) - (\gamma + \bar\eta)^{2}}}{1 + \bar\eta^{2}},
\end{equation}
which has to be positive, and
\begin{eqnarray}
   h\cos(2\bar{\theta}_{\pm})& = & 2\bar{a}^{2}_{\pm}-1,\nonumber\\
   h\sin(2\bar{\theta}_{\pm})& = & \gamma + \bar\eta (2\bar{a}^{2}_{\pm}).
\end{eqnarray}
For $\gamma\bar\eta<1$ both solutions exist, and a saddle-node
bifurcation---obtained by setting the square-root in
Eq.~(\ref{eq:AMPuniform}) to zero---occurs at
\begin{equation}\label{SNuniform}
   h_{sn}(\bar\eta)=\frac{\gamma+\bar\eta}{\sqrt{1+\bar\eta^{2}}},
   \quad \textrm{where} \quad \bar\eta= \frac{\eta}{2}.
\end{equation}
For $\gamma\bar\eta>1$ the bifurcation from the zero solution becomes
supercritical, occurring at the instability boundary of the zero
solution $h=\sqrt{1+\gamma^2}$. Note that apart from rescaling $\eta$
by a factor of $3/4$ and $a_\pm$ by a factor of $\sqrt2$ these
expressions are identical to those for the approximate amplitude and
phase of the soliton solutions~(\ref{aprx_amp}-\ref{sn_g}).

The modulational instability of the uniform solutions can be
evaluated~\cite{barashenkov03} by adding perturbations of the form
$\exp(\pm ikX)$ and calculating their growth rates using the
eigenvalues of the matrix $J^{-1}H$, obtained from
Eqs.~(\ref{H_matrix_def}) and~(\ref{H}) by substituting
$-\partial^{2}_{X}=k^{2}$, $R=\bar{a}_{\pm}\cos\bar{\theta}_{\pm}$,
and $I=-\bar{a}_{\pm}\sin\bar{\theta}_{\pm}$. The uniform solutions
are stable against such modulations as long as the larger of the real
parts of the two eigenvalues of $J^{-1}H$ is not positive for any real
$k$.  This requirement translates to satisfying the inequality
\begin{equation}\label{cond1}
  s^2 + 2(1 - 4\bar{a}_{\pm}^{2})s
  \pm 8\bar{a}_{\pm}^{2}\sqrt{h^{2}(1 + \bar\eta^{2}) - (\gamma+
  \bar\eta)^{2}}\geq0 
\end{equation}
for any non-negative $s=k^2$, where a positive sign is assumed for the
square root which is real for $h\geq h_{sn}(\bar\eta)$. 

\begin{figure}
  \includegraphics[width=0.95\columnwidth]{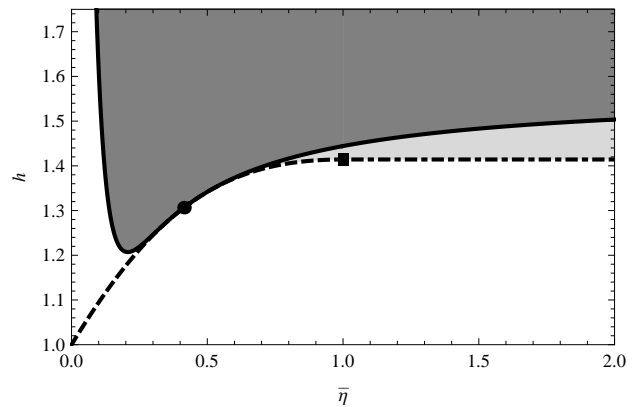}%
  \caption{\label{threshold}Stability boundary of the
    large-amplitude uniform solution $\bar\psi_+$ for $\gamma=1$. The
    solution exists above the dashed and dot-dashed curves. The dashed
    curve is the saddle-node $h_{sn}(\bar\eta)$
    [Eq.~(\ref{SNuniform})], which is replaced at the point marked
    with a small square by the dot-dashed curve, indicating the
    supercritical bifurcation from the zero solution at
    $h=\sqrt{1+\gamma^2}=\sqrt2$.  The solid curve shows
    $h_{th}(\bar\eta)$ [Eq.~(\ref{eq:htheshold})] above which ${\cal
      D}<0$. It coincides with the dashed curve at $\bar\eta_c=-\gamma
    + \sqrt{1+\gamma^2}=\sqrt2 - 1$, indicated by a small circle.  In
    the dark-gray region $\bar\psi_+$ is stable because both zeros of
    the quadratic function in~(\ref{cond1}) are complex. In the
    light-gray region $\bar\psi_+$ is stable because both zeros are
    real and negative.  This implies that for $\bar\eta\geq\bar\eta_c$
    the large-amplitude solution is always stable, and for
    $\bar\eta<\bar\eta_c$ the drive $h$ has to exceed the threshold
    value $h_{th}(\bar\eta)$ [Eq.~(\ref{eq:htheshold})] before the
    solution becomes stable.  Recall that $\bar\eta=\eta/2$.}
\end{figure}

As expected, the small-amplitude uniform solution $\bar\psi_-$ is
unstable even against a uniform perturbation, as the left-hand side of
(\ref{cond1}) is negative for $k=0$. For the large-amplitude uniform
solution $\bar\psi_+$ the inequality~(\ref{cond1}) is satisfied if
either both zeros of the quadratic function of $s$ on the left-hand
side are complex, or both are real and non-positive. The first
condition is satisfied if the discriminant
\begin{equation}
  \label{discriminant}
  {\cal D} = 4 - 32\bar\eta a_+^2\left(\gamma + 2\bar\eta a_+^2\right)
\end{equation}
is negative, and the second condition is satisfied if ${\cal D}\geq 0$
and $4\bar{a}_{+}^{2}\leq1$ (because the constant term in the
quadratic function is positive). Clearly, for $\bar\eta=0$ these
stability conditions are not satisfied, and the large-amplitude
uniform solution is modulationally unstable, in agreement with known
results~\cite{wang,barashenkov03}. However, we indeed find that
$\bar{\psi}_{+}$ can be stabilized with the help of nonlinear damping,
as shown in Fig.~\ref{threshold}. If $\bar\eta\geq\bar\eta_c = -\gamma
+ \sqrt{1+\gamma^2}$ then $\bar\psi_+$ is stable everywhere. For
weaker nonlinear damping the drive $h$ must exceed a threshold value
\begin{equation}
  \label{eq:htheshold}
  \begin{split}
    h_{th}(\bar\eta)=\frac{1}{2\bar\eta}\bigg\{1 + 2\gamma^2 
      \left(1+\bar\eta^2\right) + 4\gamma\bar\eta + 5\bar\eta^2\\ 
      -2 \left(\gamma + 2\bar\eta - \gamma\bar\eta^2\right) 
      \sqrt{1+\gamma^2} \bigg\}^{1/2}, 
  \end{split}
\end{equation}
determined by substituting the expression for
$a_+^2$ [Eq.~(\ref{eq:AMPuniform})] into the
discriminant~(\ref{discriminant}), and setting ${\cal D}=0$.

If the inequality~(\ref{cond1}) is not satisfied, the uniform state is
modulationally unstable, and the modulation whose growth rate is
fastest is expected to appear. This modulation corresponds to the
minimum of the quadratic function on the left-hand side
of~(\ref{cond1}), with wave number $k_{fast}=\sqrt{4a_+^2 - 1}$.

We demonstrate the use of the stable uniform solution in the dynamical
formation of solitons in the bottom panels of Fig.~\ref{ilm creation},
showing a numerical simulation of the equations of motion~(\ref{eom2})
with periodic boundary conditions, using $N=200$ resonators. The array
is initiated with the large-amplitude uniform solution and is driven
within the stability boundary of this state. After a long time during
which the uniform solution remains stable, the drive amplitude is
lowered below the stability threshold~(\ref{eq:htheshold}) for this
solution, but within the stability boundaries of the soliton
solutions, and solitons are formed via a modulation of the unstable
uniform state. The wavenumber of the modulation that is observed
numerically agrees with the predicted value $k_{fast}$ to within
rounding to the nearest mode satisfying the periodic boundary
conditions.

\section{SOLITON INTERACTIONS}
\label{interactions}

After finding a family of stable soliton solutions of
Eq.~(\ref{amp_eq}), it is natural to consider the interaction between
them. Soliton interactions were studied in detail in the integrable
NLS equation, corresponding to $\gamma=h=\eta=0$ in
Eq.~(\ref{PDNLS})~\cite{gordon, desem}.  It was found that the
interaction of initially stationary solitons depends on their relative
phase, with in-phase and out-of-phase solitons attracting and
repelling each other, respectively. This property is generic and is
not predicated on the integrability of the underlying equations. It is
also valid for multidimensional equations~\cite{multiD}. In the
presence of additional effects such as amplification and damping the
soliton interaction problem is not amenable to a complete analytical
study~\cite{seva,BoundStates}.  However, it is possible to analyze the
interaction between two weakly overlapping $\Psi_{+}$ solitons by
regarding the overlapping nonlinear terms---arising from the
substitution of a two-soliton solution into the PDNLS equation---as
small perturbations~\cite{longhi96,Karpman,longhi97}. In this case as
well, in-phase solitons attract each other, whereas out-of-phase
solitons repel.  Using equations derived in Refs.~\cite{BoundStates,%
  longhi97,seva} it is easy to show that adding a small nonlinear
damping term does not induce any motion on the solitons, hence one may
expect to see the same type of phase-dependent interaction in the full
amplitude equation (\ref{amp_eq}) with $\eta>0$.

\begin{figure}
    \begin{center}
    \subfigure[]{
    \includegraphics[width=0.48\columnwidth]{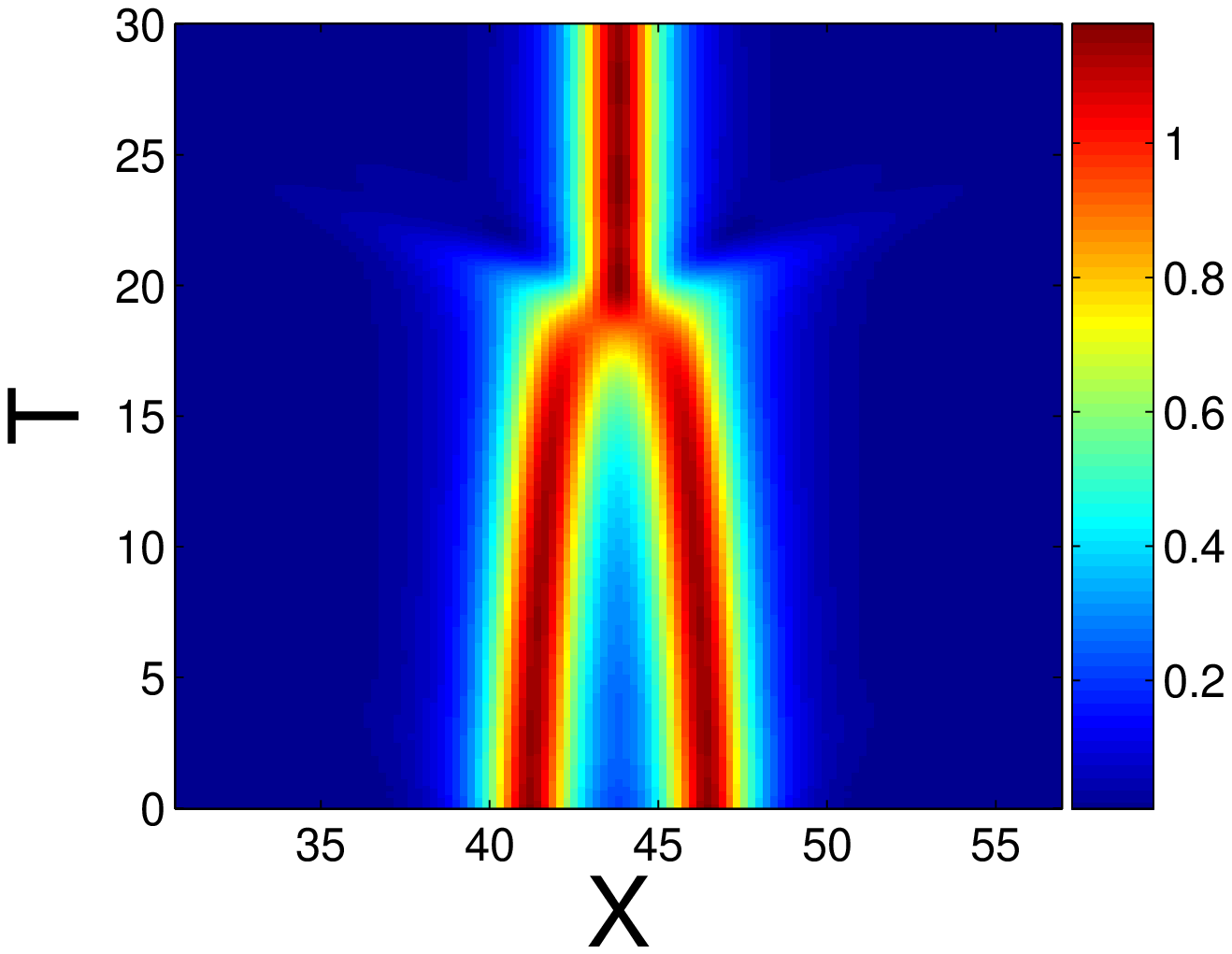}}
    \subfigure[]{
    \includegraphics[width=0.48\columnwidth]{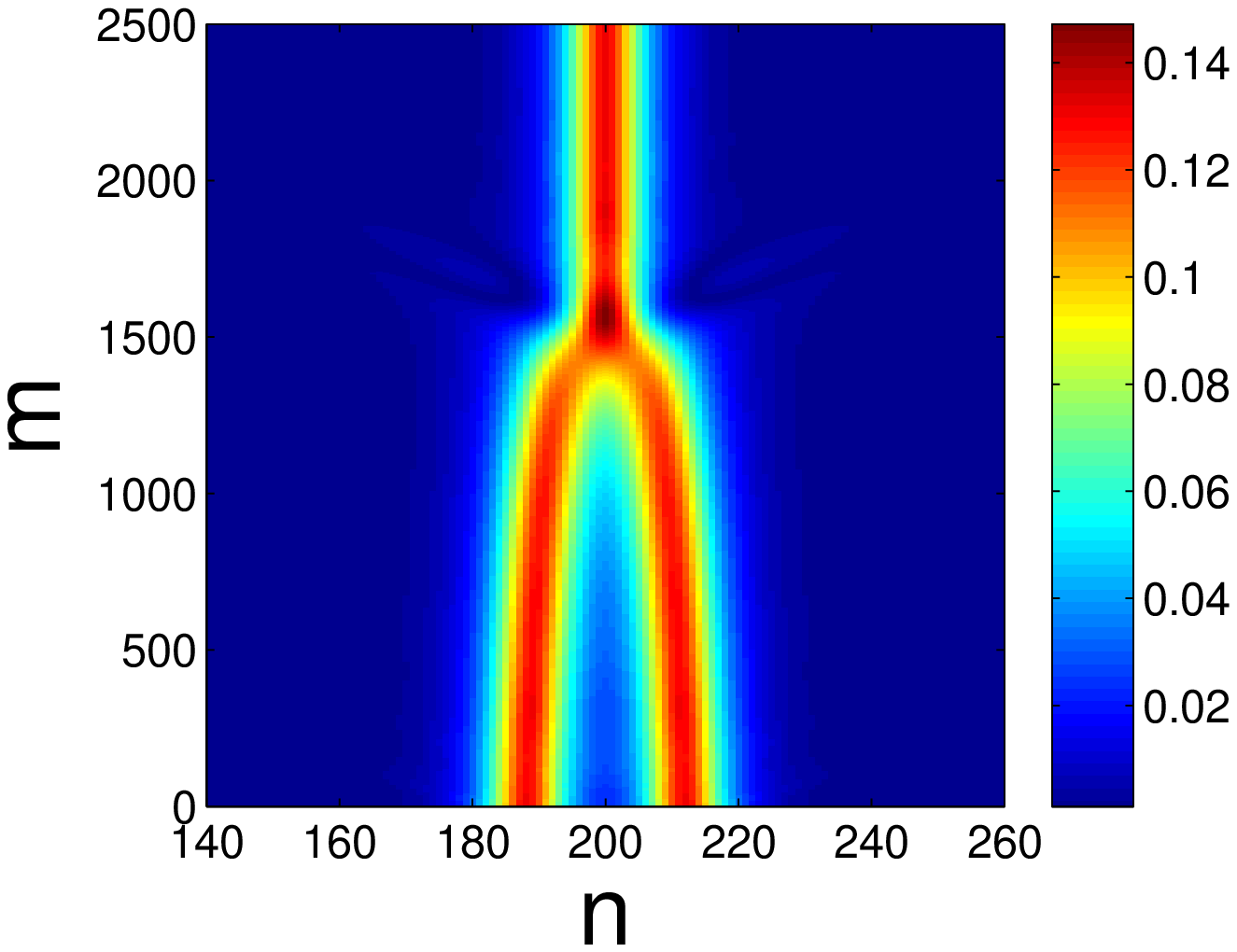}}
    \subfigure[]{
    \includegraphics[width=0.48\columnwidth]{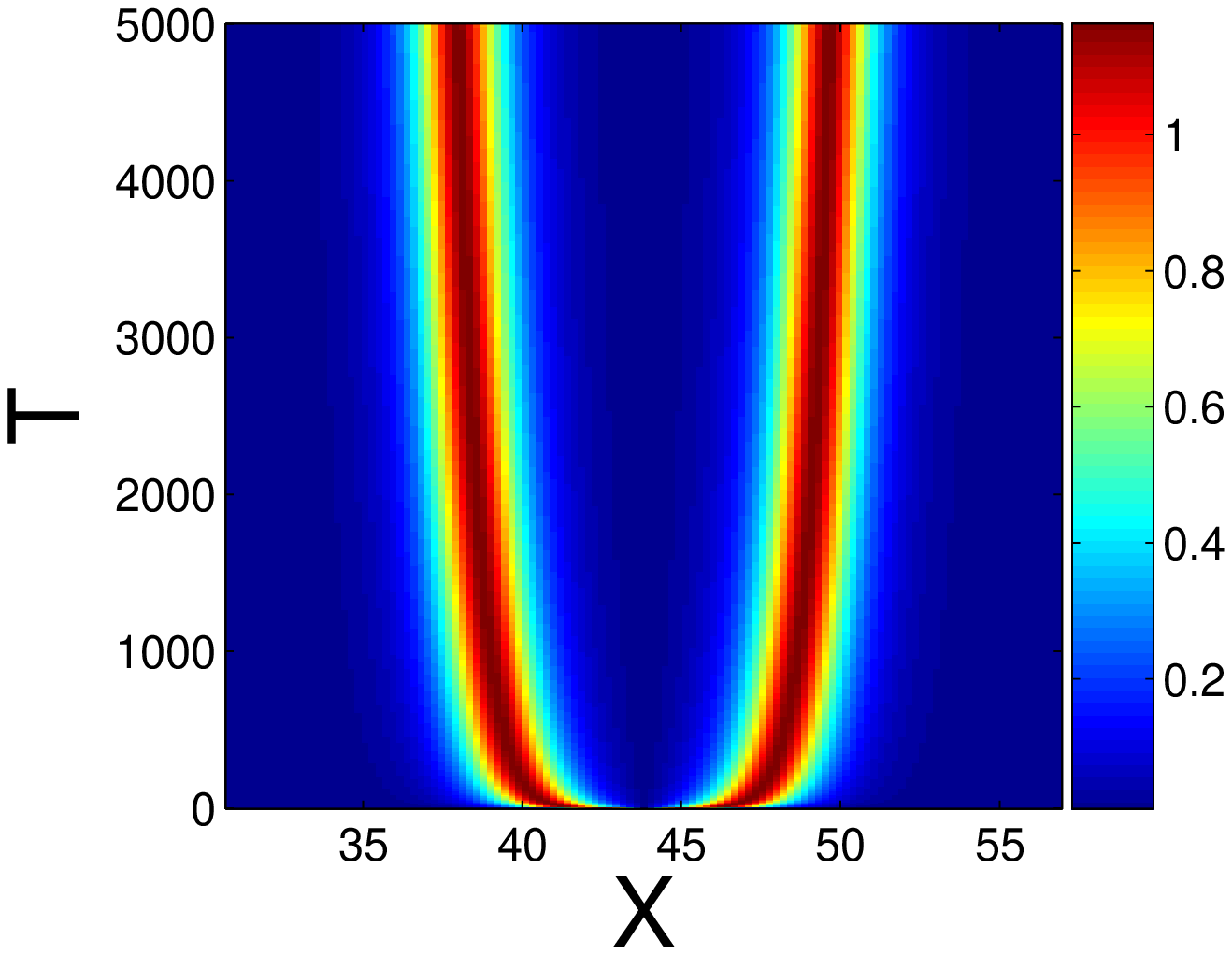}}
    \subfigure[]{
    \includegraphics[width=0.48\columnwidth]{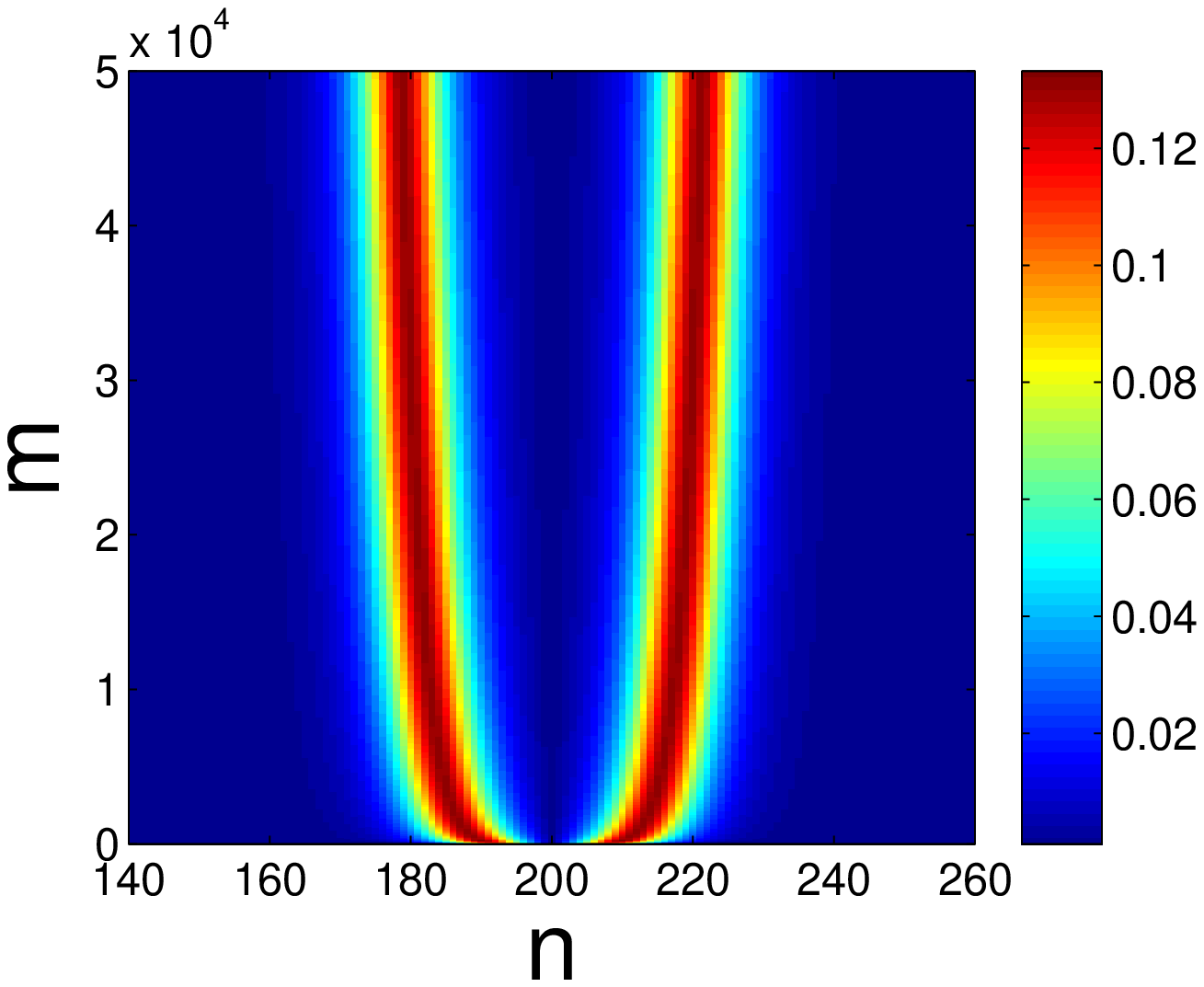}}
    \end{center}
    \caption{\label{soliton_interaction1}(Color online) Numerical
      simulations of soliton interaction. Left and right panels
      display results obtained, respectively, from numerical
      simulations of the amplitude equation~(\ref{amp_eq}) and of the
      underlying equations of motion~(\ref{eom2}), with $h=0.7$ and
      the remaining parameters as in Fig.~\ref{sweep_inset}. (a) and
      (b): Attraction between two in-phase solitons and their merger
      into a single one, after half the energy has been dissipated.
      (c) and (d): Repulsion between out-of-phase solitons.}
\end{figure}

This is indeed verified, as shown in Fig.~\ref{soliton_interaction1},
which presents the results of a numerical integration of the equations
of motion~(\ref{eom2}), and of the amplitude
equation~(\ref{amp_eq}), simulated as a PDE with initial conditions
\begin{equation}
  \label{two_solitons}
  \psi_{\pm} = \pap\left(X - (1 - r) \frac{L}{2}\right)
  \pm \pap\left(X - (1 + r)\frac{L}{2}\right),
\end{equation}
where $L=(N+1)(2\omega \Omega \epsilon/D)^{1/2}$ is the scaled length
of an array of $N$ resonators, and $rL$ is the distance between the
centers of the solitons. Note that one time unit $T=1$ of the
amplitude equation is equal to $\omega_{p}/(2\pi(\omega_{p} -
\omega))$ periods of parametric oscillations in the original equations
of motion. For the parameters used throughout this paper
$\omega_{p}/(2\pi(\omega_{p} - \omega)) \simeq 80$, and one can verify
that this is approximately the ratio between the vertical axes of the
solutions of the amplitude equation
[Fig.~\ref{soliton_interaction1}(a) and (c)] and those of the
equations of motion [Fig.~\ref{soliton_interaction1}(b) and (d)]. Also
note that the ratio between the peak heights of the soliton solutions
of the amplitude equation and those of the equations of motion in
Fig.~\ref{soliton_interaction1} is approximately $\sqrt\epsilon=0.1$,
as expected from Eq.~$(\ref{u_expansion})$.

\begin{figure}
    \centering
    \subfigure[\ $r_{c}=0.021$]{
    \includegraphics[width=0.48\columnwidth]{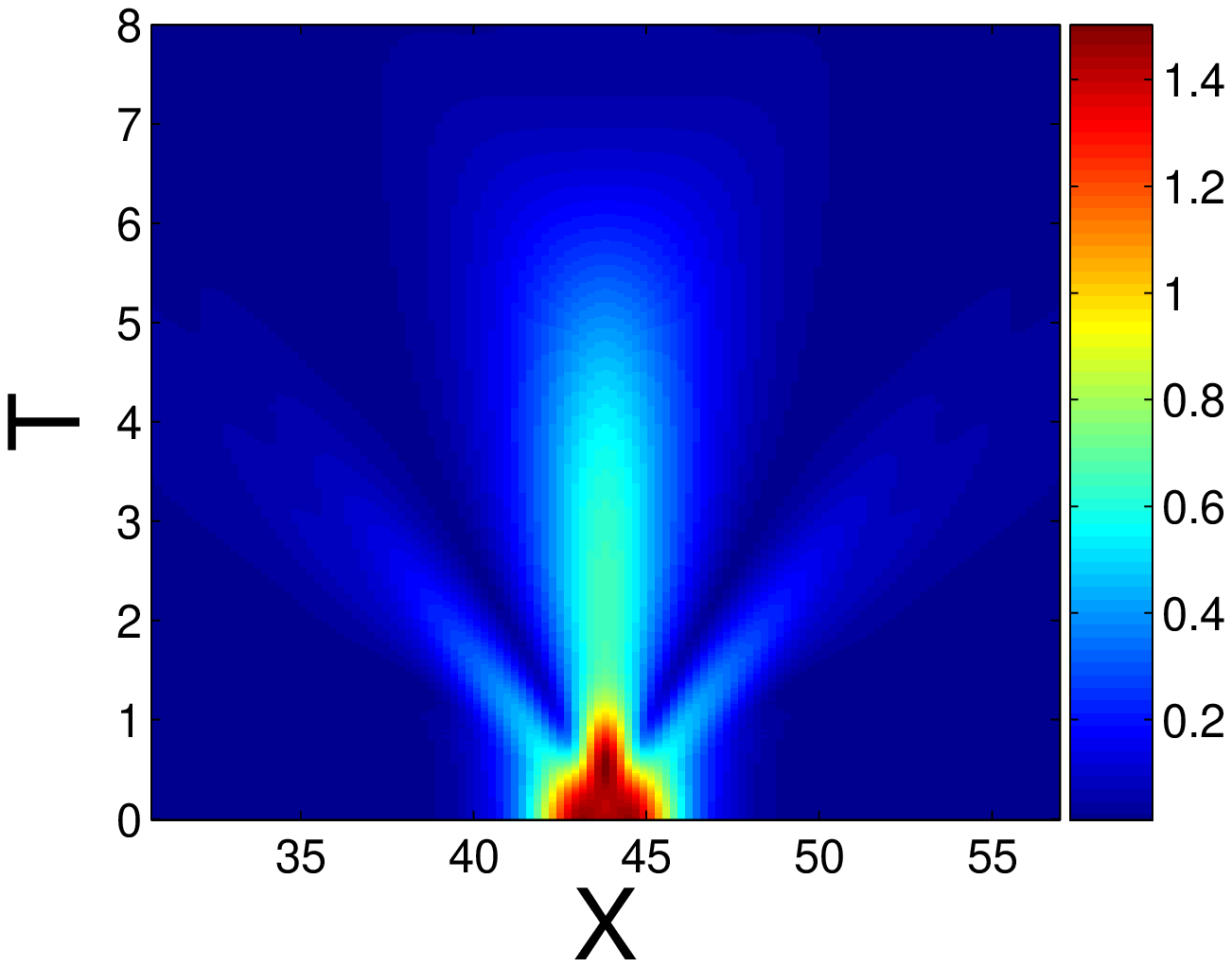}}
    \subfigure[\ $r_{c}=0.022$]{
    \includegraphics[width=0.48\columnwidth]{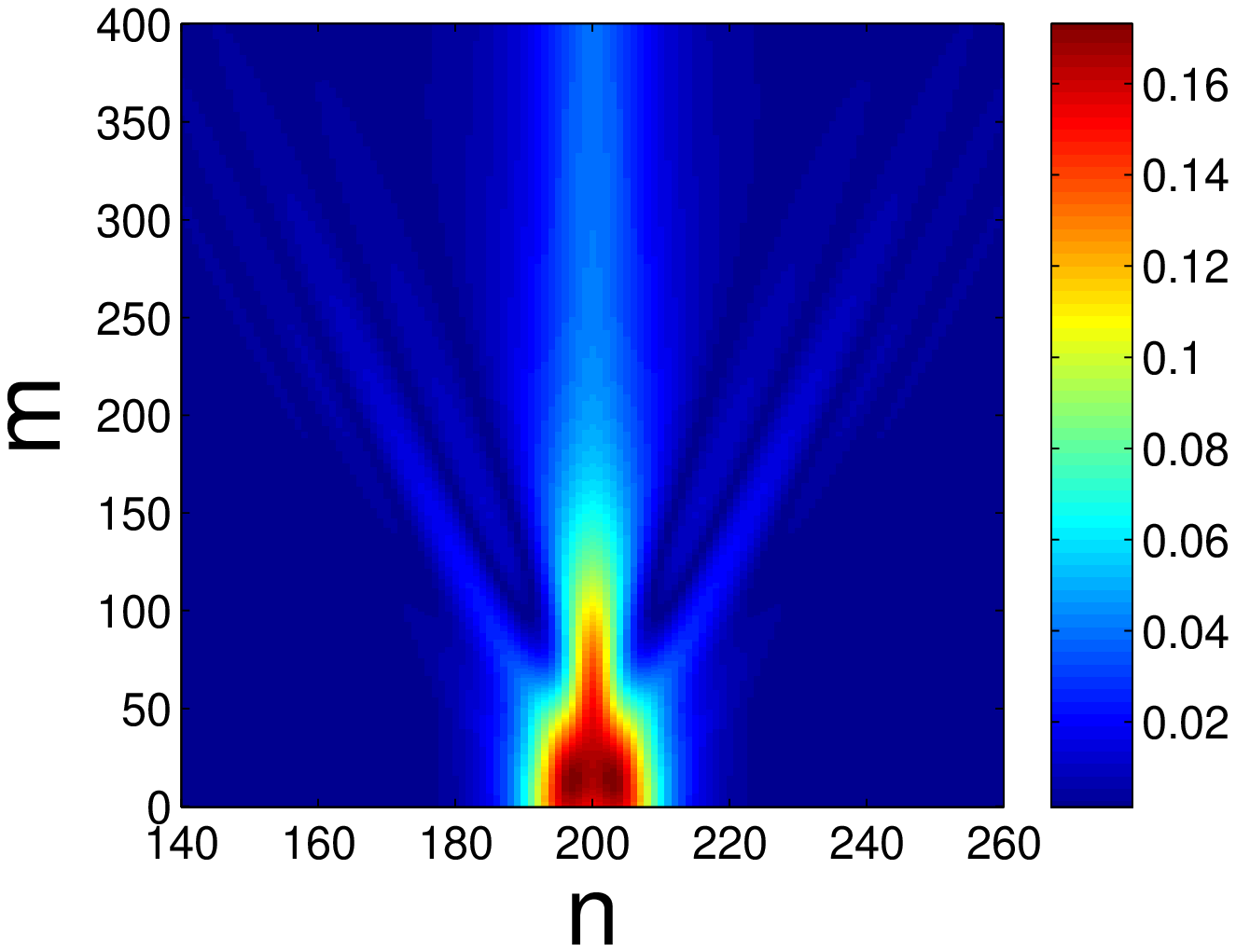}}
    \subfigure[\ $r_{c}=0.021$]{
    \includegraphics[width=0.48\columnwidth]{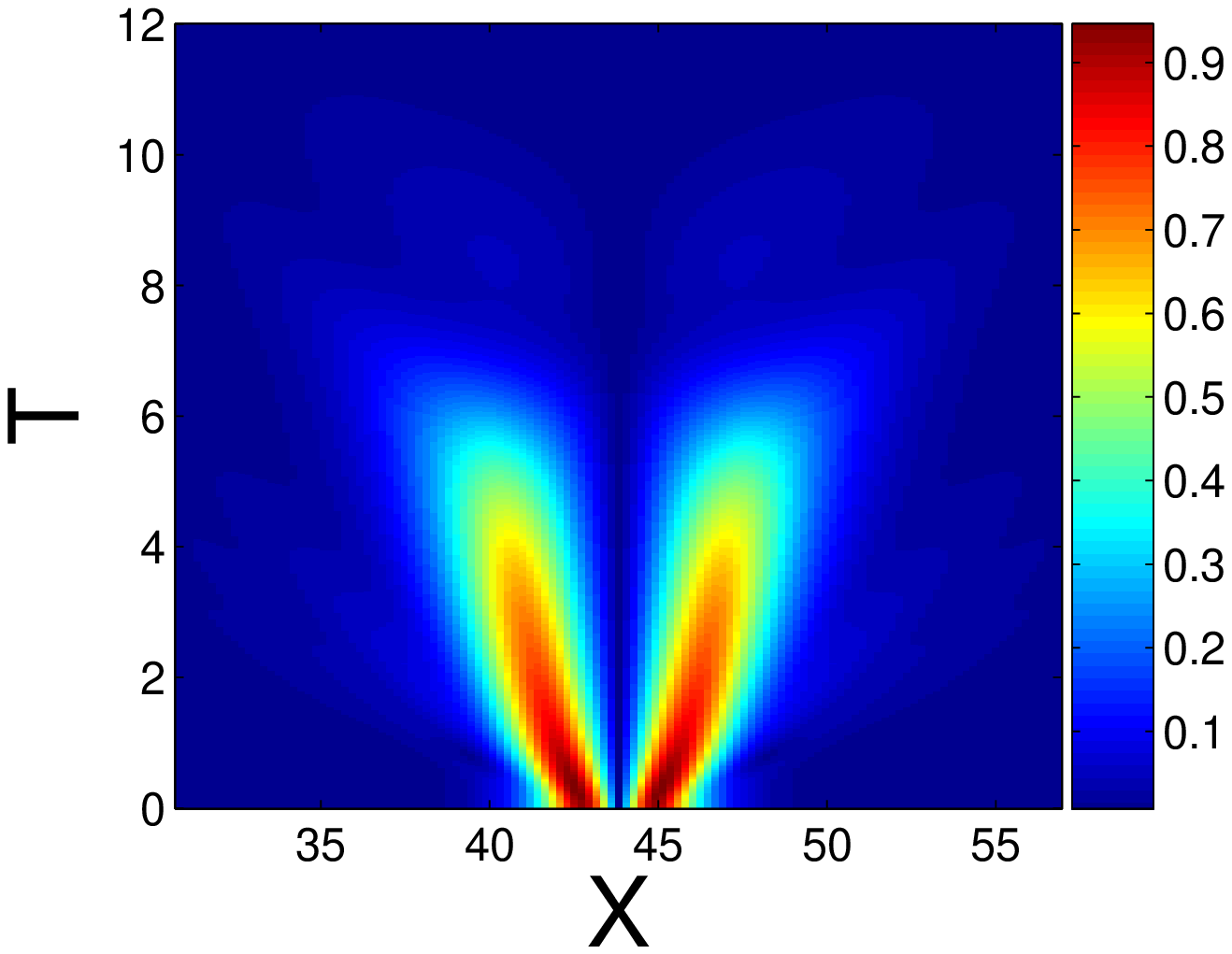}}
    \subfigure[\ $r_{c}=0.02085$]{
    \includegraphics[width=0.48\columnwidth]{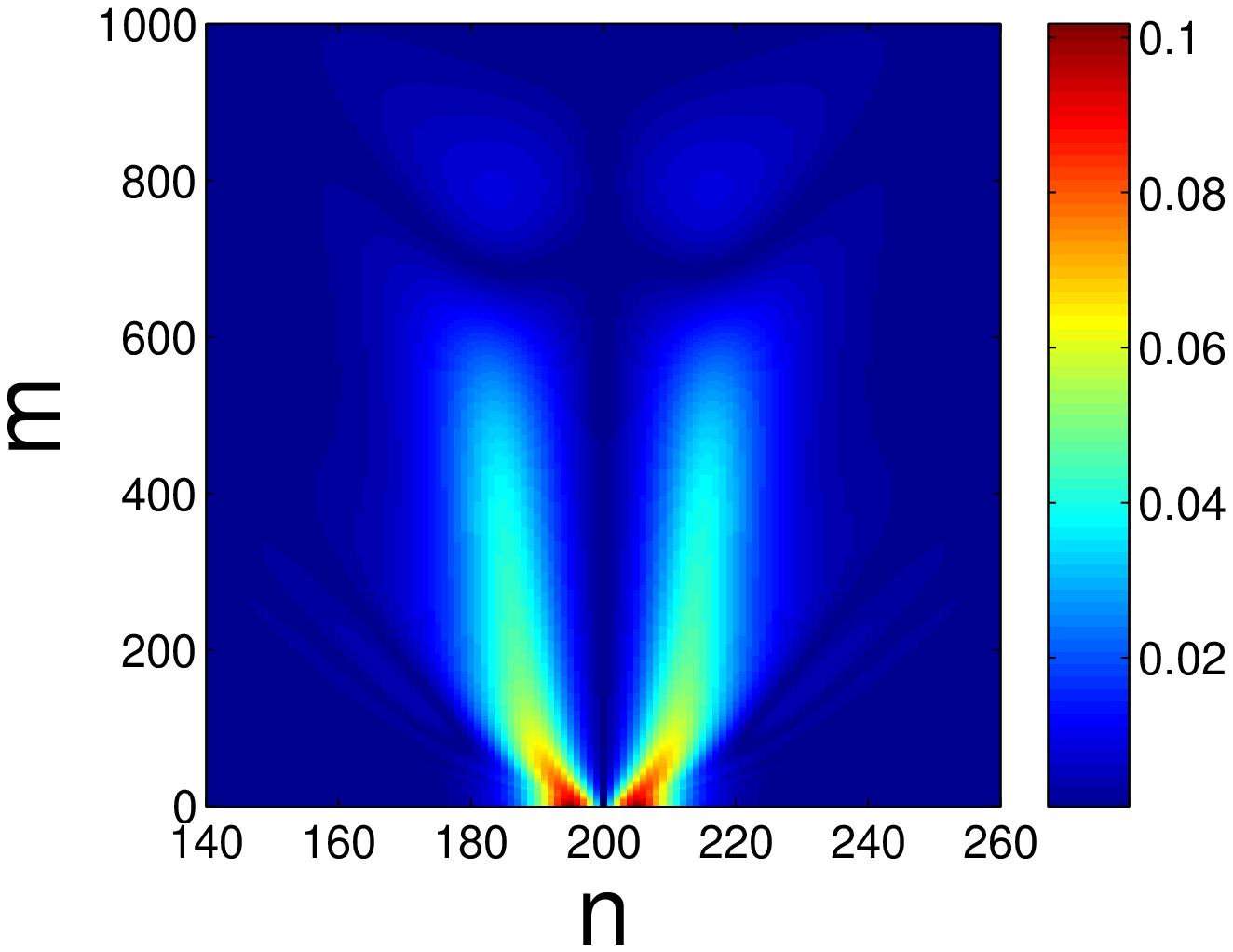}}
  \caption{\label{soliton_interaction2}(Color online) Annihilation of
    a pair of strongly overlapping solitons. Left and right panels
    display results obtained, respectively, from numerical simulations
    of the amplitude equation~(\ref{amp_eq}) and of the underlying
    equations of motion~(\ref{eom2}), initiated with a separation
    given by the indicated values of $r_c$ above which annihilation
    was not observed. (a) and (b): An initial attraction followed by
    the annihilation of a pair of in-phase solitons. (c) and (d): An
    initial repulsion followed by the annihilation of a pair of
    out-of-phase solitons. Parameters are the same as in
    Fig~\ref{soliton_interaction1}.}
\end{figure}

Another effect, which is demonstrated in
Fig.~\ref{soliton_interaction2}, is that if the solitons strongly
overlap, and $r$ is smaller than some critical distance $r_{c}$, they
annihilate into the zero state, for both the in-phase and out-of-phase
pairs. For the parameters of Fig.~\ref{soliton_interaction2}, $r_{c}
\simeq 0.021$. The annihilation of the out-of-phase pair is easily
understood because the strongly overlapping solitons cancel each
other. However, the mutual destruction of the in-phase pair is a less
obvious effect, indicating that for some reason the initial conditions
for $r< r_{c}$ are in the basin of attraction of the zero solution,
and not in that of the stable single soliton solution, contrary to the
case of $r> r_{c}$.

\section{SPLITTING SOLITONS}
\label{splitting}

\begin{figure}
    \begin{center}
    \centering
    \subfigure[\ $k=1.35$]{
    \includegraphics[width=0.48\columnwidth]{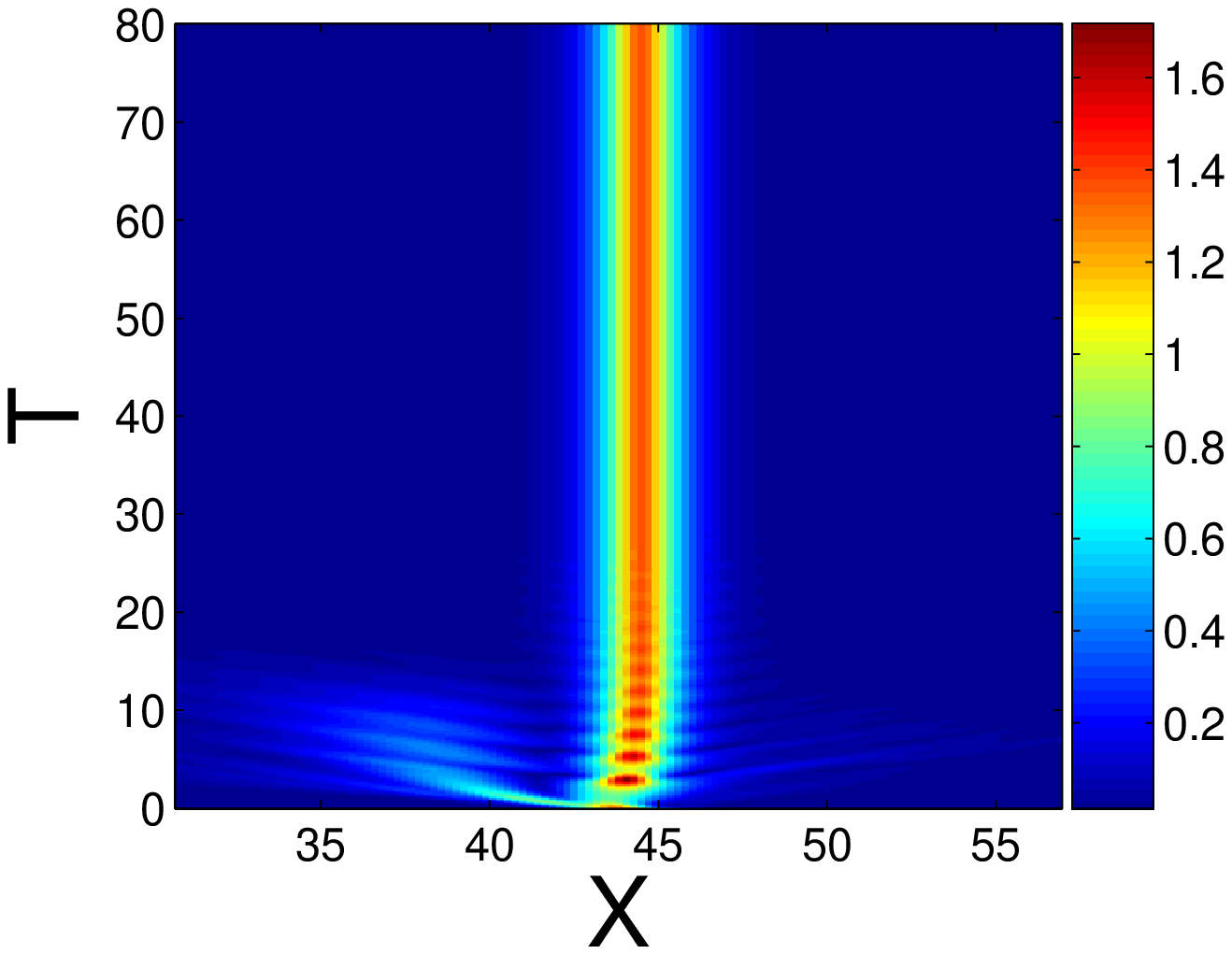}}
    \subfigure[\ $k=1.35$]{
    \includegraphics[width=0.48\columnwidth]{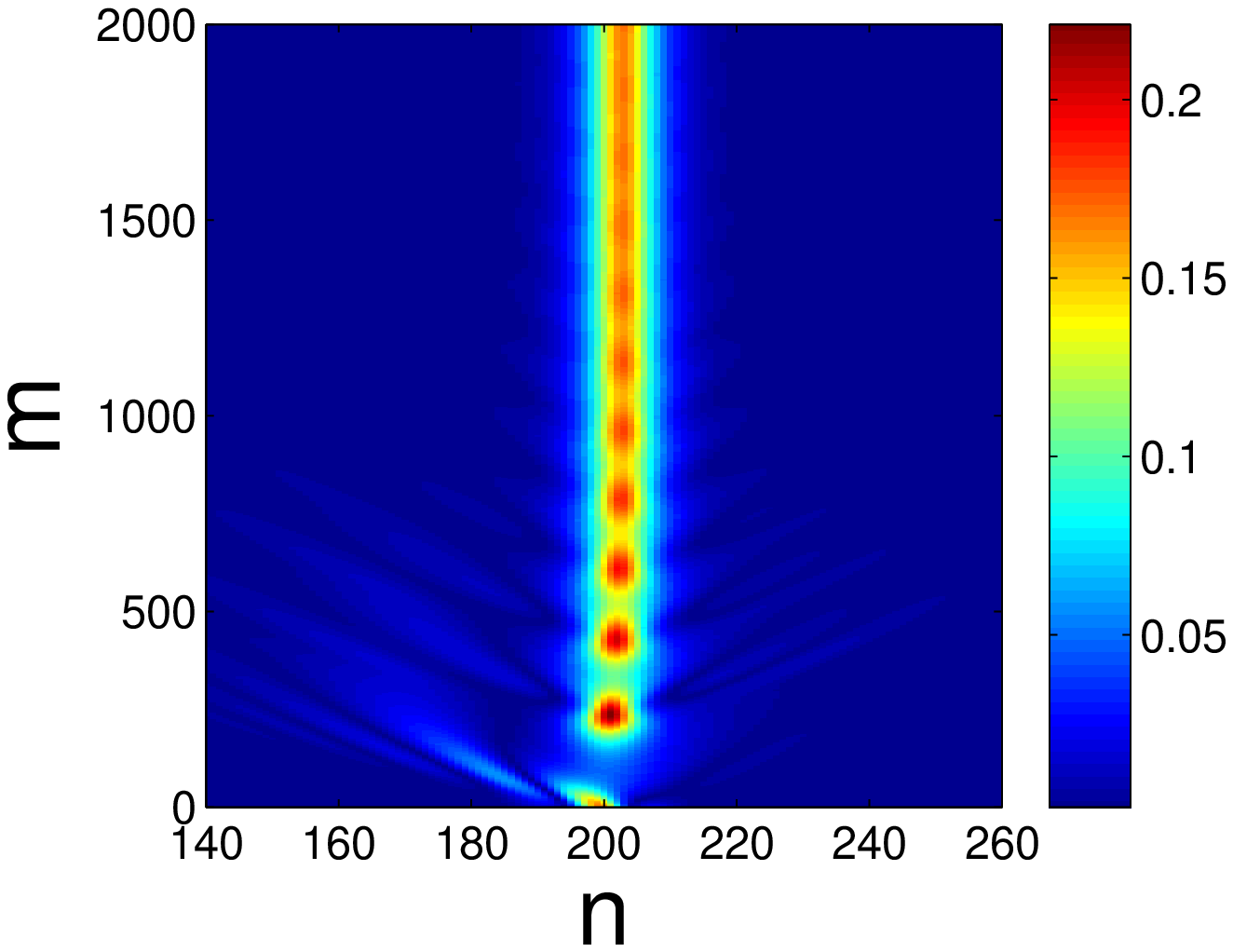}}
    \subfigure[\ $k=1.37$]{
    \includegraphics[width=0.48\columnwidth]{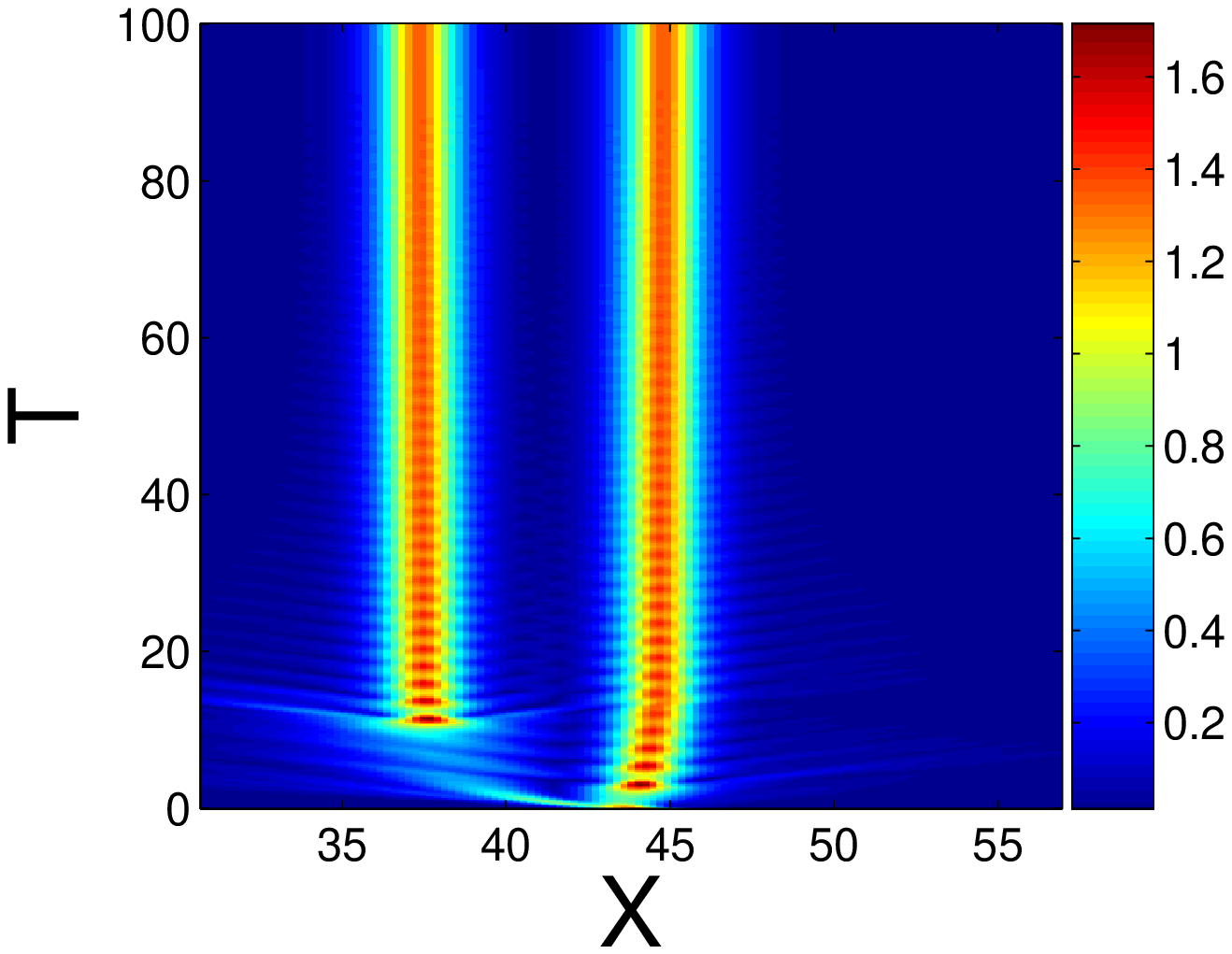}}
    \subfigure[\ $k=1.37$]{
    \includegraphics[width=0.48\columnwidth]{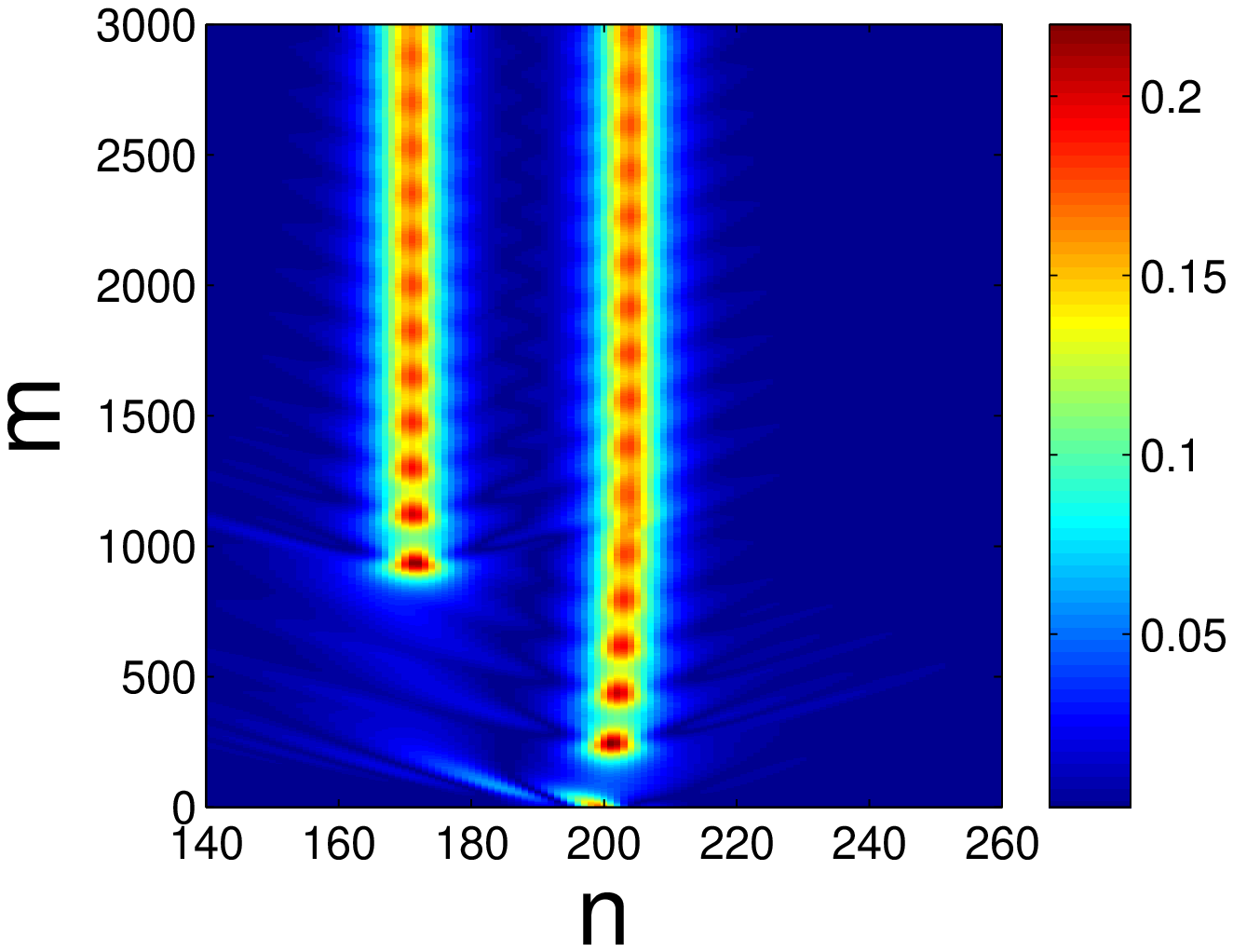}}
    \end{center}
    \caption{\label{splitting_solitons}(Color online) Simulations
      initiated with a boosted soliton. Left and right panels display
      results obtained, respectively, from numerical simulations of
      the amplitude equation~(\ref{amp_eq}) and of the underlying
      equations of motion~(\ref{eom2}), with $\eta=0.02$, $h=1$, and
      all other parameters as in Fig.~\ref{sweep_inset}. Note that the
      simulations of the equations of motion, displayed on the right,
      show only the initial stage of the evolution that is simulated
      with the amplitude equation and displayed on the left (with
      $T=1$ on the left equivalent to about $m=80$ drive periods on
      the right, as discussed in the text). (a) and (b): $k < k_{th}
      \simeq 1.36$ and the soliton moves slightly and stops.  (c) and
      (d): $k > k_{th}$ and the soliton splits into two.  The
      solutions eventually settle into the known states of one or two
      stationary solitons.}
\end{figure}

The Galilean invariance of the NLS equation admits the motion of any
solution at a constant velocity. The parametric drive $h\psi^\ast$
breaks this property of the equation.  Nevertheless, stable traveling
solitons in the parametrically-driven (but undamped) NLS equation were
obtained in a numerical form by Barashenkov \emph{et
  al.}~\cite{barashenkov01}. We have attempted to do the same with
solutions of the full amplitude equation~(\ref{amp_eq}) by multiplying
the approximate solution $\pap$ by $e^{-ik(X-X_{0})}$, thereby
\emph{boosting} it. We have concluded that such a boost may set the
soliton into transient motion, but eventually it comes to a complete
halt, as shown in Figs.~\ref{splitting_solitons}(a) and (b).

For certain parameter values we observe a noteworthy effect in which a
boosted soliton splits into two.  In order to estimate the threshold
value $k_{th}$ for the wavenumber $k$, above which this splitting
occurs, we write the energy of the soliton using the Hamiltonian
density that gives rise to the driven, but undamped, NLS equation,
\begin{equation}
  \label{energy}
  E = \int_{-\infty}^{\infty} \left(\left|\frac{\partial\psi}{\partial
        X}\right|^{2} + |\psi|^{2} - |\psi|^{4} +
    h\textrm{Re}(\psi^{2})\right)dX.
\end{equation}
Following Eq.~(\ref{amp_eq}), this energy evolves in time according to
\begin{eqnarray}
  \label{energy rate}\nonumber
  \frac{dE}{dT} &= &2\int_{-\infty}^{\infty} \left(\gamma +
    \eta|\psi|^{2}\right)\bigg(\frac{1}{2} \left(\psi
    \frac{\partial^{2}\psi^{*}}{\partial X^{2}} + \psi^{*}
    \frac{\partial^{2}\psi}{\partial X^{2}}\right)\\ &-
  &|\psi|^{2} + 2|\psi|^{4} - h\textrm{Re}(\psi^{2})\bigg)dX.
\end{eqnarray}
The right hand side of (\ref{energy rate}) is zero for
$\psi=ae^{-i\theta}\textrm{sech}(aX)$ and $h\cos(2\theta)=a^{2}-1$ for
any constant $a$---in particular, for the approximate solution $\pap$
as well as the numerically exact solution.  By substituting the boosted
approximate solution $\pap(X) e^{-ik(X-X_{0})}$ into~(\ref{energy}), we
find its energy to be
\begin{equation}
  \label{energy_boost}
  E(k) = 2a_{+}(1 + k^{2}) + \frac{2k\pi(a_{+}^{2} -
    1)}{\sinh(k\pi/a_{+})} - \frac{2}{3}a_{+}^{3},
\end{equation}
whereas for the static soliton $E(k=0) = 4 a_{+}^{3}/3$. Thus, an
obvious estimate for the threshold wavenumber $k_{th}$ required to
split a soliton into two is given by the condition
$E(k_{th})=2E(0)$. For the parameters of
Fig.~\ref{splitting_solitons}, $k_{th}\simeq1.36$, and indeed below
this value the soliton does not split [in (a) and (b) $k=1.35$] while
above this value the soliton does split [in (c) and (d) $k=1.37$]. At
still larger values of $k$ the soliton is destroyed by the boost and
eventually decays to zero. For the parameters of
Fig.~\ref{splitting_solitons} this happens for $k>1.59$. We note that
although it might seem plausible to have values of $k$ for which
boosting a single soliton would split it into three, we were unable to
detect such an effect.

\section{BOUND STATES}
\label{complexes}

\begin{figure}[t]
\includegraphics[width=0.9\columnwidth]{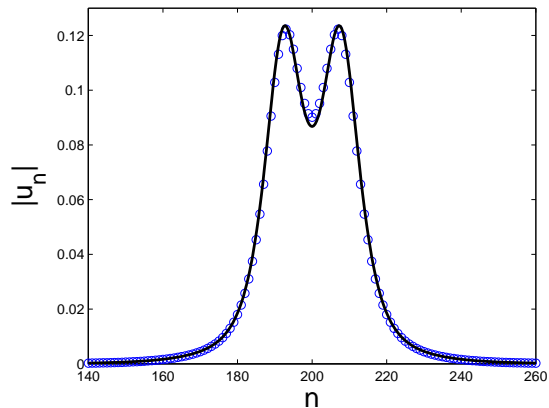}%
\caption{\label{complex}(Color online) A stable bound state of two
  solitons. The $\circ$s are absolute values of the displacements of
  the resonators, obtained by a numerical integration of the discrete
  equations of motion~(\ref{eom2}), after a sufficiently long
  transient time has elapsed. The solid line is the stable solution
  obtained by solving the amplitude equation~(\ref{amp_eq}) as a
  boundary value problem.  The parameters are $\gamma=0.565, h=1,
  \eta=0.3$, and all others as in Fig.~\ref{sweep_inset}.}
\end{figure}

We have considered the effects of pairwise interaction between
solitons, and of boosting a single static soliton. It was shown by
Barashenkov and Zemlyanaya~\cite{barashenkov99} that a combination of
both features within the framework of the PDNLS equation may lead to
the formation of solitonic complexes, or bound states~\cite{%
  BoundStates}.  These complexes were found numerically, solving the
PDNLS with an initial guess of the form
\begin{equation}
  \label{anzats}
  \psi_{\rm b} = \psi(X - X_{0})e^{ik(X - X_{0})}+\psi(X +
  X_{0})e^{-ik(X + X_{0})},
\end{equation}
where $\psi(X) = A \textrm{sech}(AX)e^{-i\theta}$. For $\gamma=0.565$ and
$h=0.9$ the rest of the parameters were found by means of a variational
procedure elaborated in~\cite{barashenkov99} to be
$\theta=\Theta_{+}$, $A=1.14$, $k=-0.068$, and $X_{0}=2.017$.

Using the PDNLS variational ansatz~(\ref{anzats}) as an initial guess,
we are able to obtain stationary solitonic bound states for the full
amplitude equation~(\ref{amp_eq}) with $\eta>0$.  Performing a linear
stability analysis on these solutions, as described earlier using
Eq.~(\ref{H_matrix_complex}), reveals that some of them are stable.
Fig.~\ref{complex} shows one of these stable bound state solutions,
obtained numerically using the amplitude equation~(\ref{amp_eq}), and
nicely reproduced by a numerical integration of the underlying
equations of motion~(\ref{eom2}).

\section{CONCLUSIONS}
\label{sec:conclusions}

We have investigated intrinsic localization of vibration in response
to parametric excitation in an array of resonators with a
\emph{stiffening} nonlinearity. Our analysis was chiefly performed on
a single amplitude equation, which was derived directly from the
underlying equations of motion of the array to describe the slow
spatio-temporal dynamics of the system.  The discreteness of the array
imposes an upper bound on the spectrum of linear modes. We have
studied the case in which neighboring resonators oscillate
out-of-phase, in the staggered mode, with an oscillation frequency set
slightly above the top frequency of the linear spectrum. One can
similarly study ILMs in resonators with a \emph{softening}
nonlinearity, by changing the sign of $u_n^3$ in the equations of
motion~(\ref{eom2}), and considering the case in which neighboring
resonators oscillate in-phase with an oscillation frequency set
slightly below the bottom frequency of the linear spectrum.

The array that we consider, hence also the amplitude equation we
derive, are nonlinearly damped. Its localized modes emerge from two
exact soliton solutions that exist in the absence of nonlinear
damping. We have shown that nonlinear damping increases the range of
parameters for which localized solutions are stable. However,
nonlinear damping increases the region in which the zero state is the
only stable one, and it also stabilizes the non-zero uniform solution
of the amplitude equation, which is modulationally-unstable if
$\eta=0$.  We have studied soliton interaction and soliton splitting,
both in the presence of nonlinear damping. We have also found a family
of localized solutions in the form of bound states of two solitons. In
a follow-up work, we intend to perform a more detailed investigation
of the different localized solutions of the full amplitude
equation~(\ref{amp_eq}) with $\eta>0$, using numerical continuation.

All results obtained from the amplitude equation are in excellent
agreement with numerical solutions of the underlying equations of
motion. This upholds the validity of using a continuous PDE as a tool
for analyzing ILMs, or discrete solitons, in a system whose original
description is given in terms of coupled ordinary differential
equations.  Furthermore, our numerical simulations of the equations of
motion suggest that the predicted effects can be observed in
parametrically-driven arrays of real MEMS and NEMS resonators, thus
motivating new experiments in these systems.

We thank an anonymous referee for inquiring about the dynamical
formation of solitons in our array, which prompted us to add
Sec.~\ref{flat}. This work was supported by the U.S.-Israel Binational
Science Foundation (BSF) through Grant No.~2004339, and by the Israeli
Ministry of Science and Technology.

\bibliography{ILM}
\end{document}